\documentclass{aa}  

\usepackage[varg]{txfonts}
\usepackage{microtype}

\usepackage{natbib}
\bibpunct{(}{)}{;}{a}{}{,}

\usepackage{placeins}
\usepackage{capt-of}

\usepackage{graphicx}

\usepackage{units}
\usepackage[breaklinks=true]{hyperref}

\makeatletter 
\renewcommand*\aa@pageof{, page \thepage{} of \pageref*{LastPage}}
\makeatother

\begin{document}

   \title{Dust destruction by the supernova remnant forward shock\\in a turbulent interstellar medium}

   \author{Tassilo Scheffler \inst{\ref{inst1}} \href{http://orcid.org/0009-0002-3600-4516}{\includegraphics[width=9pt]{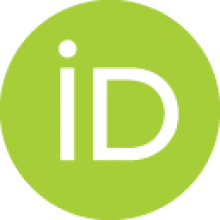}}
          \and
          Nina S. Sartorio
          \inst{\ref{inst1}}\href{https://orcid.org/0000-0003-2138-5192}{\includegraphics[width=9pt]{logo_orcid.png}}
          \and
          Florian Kirchschlager
          \inst{\ref{inst1}}\href{https://orcid.org/0000-0002-3036-0184}{\includegraphics[width=9pt]{logo_orcid.png}}
          \and\\
          Ilse De Looze
          \inst{\ref{inst1}}\href{https://orcid.org/0000-0001-9419-6355}{\includegraphics[width=9pt]{logo_orcid.png}}
          \and
          Michael J. Barlow
          \inst{\ref{inst2}}\href{https://orcid.org/0000-0002-3875-1171}{\includegraphics[width=9pt]{logo_orcid.png}}
          \and
          Franziska D. Schmidt
          \inst{\ref{inst2}}
          }

   \institute{Sterrenkundig Observatorium, Ghent University, Krĳgslaan 299, B-9000 Ghent, Belgium\label{inst1}\\
        \email{\href{mailto:tassilo.scheffler@ugent.be}{tassilo.scheffler@ugent.be}}
        \and
        Department of Physics and Astronomy, University College London, Gower Street, London WC1E 6BT, UK\label{inst2}
             }

   \date{Received 18 July 2025 / Accepted 30 November 2025}

   \authorrunning{Tassilo Scheffler et al.}
 
  \abstract
  {While supernova remnants (SNRs) have been observed to produce up to \unit[1]{M$_\odot$} of dust, the amount of dust destroyed by the forward shock is poorly constrained, raising the question of whether they are net dust producers or destroyers.}
  {Our aim was to estimate the dust destruction efficiency of SNR forward shocks in a realistically turbulent interstellar medium (ISM) during their most destructive phase, and to assess dust shielding by high-density filaments during this period.}
  {We ran 3D high-resolution turbulence simulations for different turbulent Mach numbers (0--3) and average ISM densities (\unit[1--100]{cm$^{-3}$}) to resemble observations of the turbulent ISM. We then set off a supernova explosion to trace its 3D magnetohydrodynamic evolution for the first \unit[10]{kyr}. Finally, we ran post-processing simulations to investigate the dust transport and destruction by the SNR forward shock, taking into account gas and plasma drag, kinetic and thermal sputtering, and grain-grain collisions, and considering either silicates or carbonaceous dust.}
  {The dust destruction rate of the forward shock strongly depends on the average ISM density and turbulence strength, varying in the range 27--92\% (\unit[0.85--11.0]{M$_\odot$}) in the studied \unit[10]{kyr}. Overall, dust is less efficiently destroyed in a low-density medium ($\unit[1]{cm^{-3}}$, 27--57\%) than in intermediate-density ($\unit[10]{cm^{-3}}$, 46--92\%) and high-density ($\unit[100]{cm^{-3}}$, 73--87\%) cases. The forward shock is found to destroy 8--34\% less dust in high Mach turbulence compared to a homogeneous medium. Furthermore, carbonaceous grains are up to 21\% more robust than silicates.}
  {Filaments can partly shield dust from destruction in the first \unit[10]{kyr};  however, always more than \unit[0.85]{M$_\odot$} of dust is destroyed, making most SNRs dust sinks under the conditions explored in this work. The destruction efficiency of the SNRs with less than \unit[1]{M$_\odot$} of destroyed dust has not yet plateaued, so that they are most likely also net dust destroyers by the end of their lifetimes.}
  
   
   \maketitle

\begin{nolinenumbers}
\section{Introduction}\label{sec:intro}

Understanding dust formation and destruction is key for planetary, stellar, and (extra-)galactic astronomy alike \citep[cf.][]{Draine2003, Micelotta2018}. Asymptotic giant branch (AGB) stars and supernova remnants (SNRs) are known to provide favorable conditions for dust production.
Formed dust grains are also hypothesized to be able to grow efficiently in the interstellar medium \citep[ISM; e.g.,][]{Draine2009, Hirashita2012, Mattsson2012, Zhukovska2016, Zhukovska2018}, significantly increasing the ISM dust budget over time. Even so, the relative importance of these dust formation and growth mechanisms across redshifts and the detailed underlying physical processes remain uncertain \citep[cf.][]{Schneider2024}.

Asymptotic giant branch stars play an important role in the dust budget of local galaxies \citep[e.g.,][]{Matsuura2009, Olofsson2010, Schneider2014, Dell'Agli2015}. However, at high redshifts, where several galaxies have been observed to contain high dust masses \citep[e.g.,][]{Pettini1994, Bertoldi2003, Watson2015, Bowler2022, Sommovigo2022, Akins2023, Witstok2023}, the dust needs to be produced at a much faster rate than typical AGB dust production channels can provide. AGB stars are the end evolutionary stages of low- and intermediate-mass stars with masses of less than \unit[8]{M$_\odot$}, and thus take several tens of million years to start contributing to the dust budget \citep[e.g.,][]{Vassiliadis1993, Valiante2009}. If the initial mass function is bottom-heavy then this contribution will take even longer with solar mass stars taking billions of years to start their dust production. Therefore, the majority of the stars could not aid in dust production at high redshifts. Past studies show that AGB stars in early galaxies need at least \unit[300]{Myr} to dominate the dust production \citep{DiCriscienzo2013, Valiante2017}, which is too long to account for the earliest dusty galaxies.

In recent decades, core-collapse supernovae (CCSNe) have been found to be a promising candidate for dust production in the early universe \citep[e.g.,][]{Morgan2003, Dwek2007, Gall2011b, Gall2011c, Gall2011, Lesniewska2019, Pozzi2021, Dayal2022, Palla2024} because they originate from stars more massive than \unit[8]{M$_\odot$}, which have lifetimes of only a few to several tens of million years. Observations in the local universe have shown that, within the first few decades after the SN explosion, \unit[1]{M$_\odot$} of dust can be formed in the SNR ejecta \citep[e.g.,][]{Lucy1989, Dunne2009, Matsuura2011, Gomez2012, Temim2012, Bevan2016, DeLooze2017, DeLooze2019, Niculescu-Duvaz2022}. However, SNRs do not only produce dust; they are also able to destroy a large amount of pre-existent dust, due to their high-velocity forward shock \citep[e.g.,][]{2015Lakicevic, Priestley2021b, Priestley2022}. This shock induces gas-grain collisions (sputtering) and grain-grain collisions (shattering), thus reducing the ISM dust grain sizes or destroying them completely \citep[e.g.,][]{Barlow1978, Draine1979, Jones1994, Jones1996}. Furthermore, the freshly produced ejecta dust needs to survive the reverse shock that forms in a subsample of SNRs with heavy pre-SN mass-loss or high-density environment as a result of pressure build-up behind the forward shock. This could result in a significantly lower dust contribution from SNRs to the ISM than the commonly assumed \unit[1]{M$_\odot$} \citep[e.g.,][]{Kirchschlager2024b}. When forward and reverse shock destruction are taken into account, it becomes unclear whether SNe can indeed make a net contribution to the dust budget.

To resolve this uncertainty, it is crucial to constrain the dust destruction rate of the SNR forward shock, which encounters several tens of solar masses of dust during its evolution \citep[e.g.,][]{Kirchschlager2022, Kirchschlager2024a}. 
Many analytical and numerical forward shock dust destruction studies have been carried out \citep[e.g.,][]{Jones1994, Jones1996, Nozawa2006, Bocchio2014, Slavin2015, Hu2019, Martinez-Gonzalez2019, Kirchschlager2022, Kirchschlager2024a, Vasiliev2024, Dedikov2025} and predict a wide range of outcomes between almost total survival \citep[][0.03\% destruction rate]{Martinez-Gonzalez2019} to total destruction \citep[e.g.,][up to 100\% destruction rate]{Bocchio2014}. Thus, SNRs can act as dust producers or destroyers depending on the considered forward shock study and the similarly uncertain reverse shock destruction rate \citep[e.g.,][]{Bianchi2007, Nozawa2010, Silvia2010, Silvia2012, Bocchio2016, Micelotta2016, Slavin2020, Kirchschlager2019, Kirchschlager2023, Kirchschlager2024b, Vasiliev2024, Otaki2025}.

This large disparity originates from the different initial conditions, and  also from the various ways of implementing dust into codes \citep{Slavin2024} and the considered dust transport and destruction mechanisms. In particular, grain-grain collisions were expected to only have a small impact on the dust destruction, and thus were often neglected \citep[e.g.,][]{Nozawa2006, Bianchi2007, Nozawa2010, Silvia2010, Silvia2012, Micelotta2016, Hu2019, Martinez-Gonzalez2019, Slavin2020, Vasiliev2024, Dedikov2025}. However, \citet{Slavin2015} and \citet{Kirchschlager2019} found that the redistribution of grain sizes can work synergistically with other dust destruction mechanisms, and thus lead to large amounts of  destroyed dust masses. Furthermore, a magnetic field was also often not taken into account \citep[e.g.,][]{Nozawa2006, Bianchi2007, Nozawa2010, Silvia2010, Silvia2012, Martinez-Gonzalez2019, Kirchschlager2020, Slavin2020, Kirchschlager2022, Vasiliev2024, Dedikov2025}, although dust grains are charged and thus influenced by it. It was found that the effects of the Lorentz force can significantly change the dust destruction efficiency in SNRs provided that high enough magnetic field strengths are present \citep[e.g.,][]{Shull1978, Draine1979, Fry2020, Kirchschlager2023, Kirchschlager2024a}.

In addition to the dust destruction mechanisms considered, the properties of the ambient medium surrounding the SN are expected to affect how much dust is destroyed by the forward shock.
Most works consider a simple homogeneous ISM in which the SNR forward shock evolves spherically. There are just a few studies we know of that include an inhomogeneous ISM \citep{Hu2019, Kirchschlager2022, Kirchschlager2024a, Vasiliev2024, Dedikov2025}, in which the SNR evolution becomes non-spherical because the forward shock is slowed down significantly in the dense cold filaments and can travel faster through voids. Since kinetic sputtering mainly depends on the shock velocity and thermal sputtering on the temperature, it is theorized that the dense and dust-rich filaments can shield dust from destruction. Simple small density fluctuations, as considered by \citet{Kirchschlager2022} and \citet{Vasiliev2024}, show no agreement as the former finds a slightly higher dust destruction compared to the homogeneous case and the latter does not provide such a comparison. A more comprehensive study of different density inhomogeneities of \citet{Dedikov2025} draws a clearer picture. They find that a clumpy ISM can reduce the dust destruction by up to $\sim$20\%. More complex turbulence, as studied by \citet{Hu2019} and \citet{Kirchschlager2024a}, shows the same trend. 
These studies drive their turbulence by setting off SNe every \unit[1--2]{Myr} at random locations in an initially homogeneous medium and letting the system evolve for \unit[0.4]{Gyr} or to a statistical steady state, respectively. Since the turbulence is only driven by SNe, it lacks small-scale structures that influence the SNR evolution, particularly in the early stages, and are therefore expected to affect the dust destruction by the forward shock. Additionally, other feedback mechanisms such as winds and photoionization, or gravity, which would affect the turbulent energy cascade, are neglected. Thus, although this type of turbulence ensures a purely physical origin, it often has different statistical properties compared to the observed multi-scale turbulence \citep[e.g.,][]{Federrath2010} whose main driving mechanisms are still under debate \citep[e.g.,][]{Miesch1994, MacLow2004, Brunt2009, Burkhart2019, Sartorio2021}.

Turbulence can also be driven in simulations by a forcing term which can be employed by an Ornstein-Uhlenbeck process to insert Fourier waves with random propagation directions, and at random locations and scales that go down to the Jeans length. These shocks then interact until the density and velocity probability density functions (PDFs) of the fluid quantities converge. Since small and large energy waves are injected, different kinds of modes will occur in the resulting turbulence. With this method, the initial conditions can be fine-tuned to closely resemble the observed turbulence with simulations. 

We used this approach to drive purely solenoidal turbulence in a highly resolved 3D simulation box, in order to resemble the density distribution seen in regions of the ISM, including small-scale structures. Afterward, an SN was injected into the driven inhomogeneities and evolved for \unit[10]{kyr}. During this period, the SNR forward shock is the most destructive due to its high initial kinetic and thermal energy. Furthermore, it already encounters a significant amount of dust due to its rapid expansion. After \unit[10]{kyr}, the SNR will continue expanding and destroying dust, but with a lower destruction rate. By including thermal and kinetic sputtering, grain-grain collisions, and a magnetic field, we ensured that all known relevant dust transport and destruction mechanisms were considered. In this way, we aim to improve constraints on the forward shock dust destruction efficiency during the most destructive phase of SNR evolution.

We describe the methods in greater detail in Sect.~\ref{sec:methods}, followed by Sect.~\ref{sec:results}, which shows and analyzes the results of our  magnetohydrodynamic (MHD) and dust simulations. The findings are discussed and compared to previous studies in Sect.~\ref{sec:discussion}, and the paper   concludes in Sect.~\ref{sec:conclusion}.

\section{Methods}\label{sec:methods}
For this study we performed 3D hydrodynamic simulations with the widely used quasi-Lagrangian code \texttt{Arepo} \citep{Springel2010}, which uses a moving mesh based on Voronoi tessellation. We drove turbulence using an Ornstein-Uhlenbeck process (Sect.~\ref{subsec:methods_turbulence}), and considered different average ISM densities of $\unit[1]{cm^{-3}}$ (low density), $\unit[10]{cm^{-3}}$ (intermediate density), or $\unit[100]{cm^{-3}}$ (high density) and a turbulent Mach number of 0 (homogeneous), 0.3 (subsonic), 1 (transonic), or 3 (supersonic). Afterward, an SN explosion was injected as a large amount of thermal energy, distributed over a few cells at the center of our simulation domain. The SNR evolution was followed for \unit[10]{kyr} in 3D MHD simulations with the same code (Sect.~\ref{subsec:methods_SNR}). Since \texttt{Arepo} does not include detailed dust transport and destruction mechanisms, we used the dust post-processing code \texttt{Paperboats} \citep{Kirchschlager2019} that assumes an initial dust distribution in the turbulent box and then calculates the dust transport and destruction based on gas and plasma drag, Lorentz acceleration, thermal and kinetic sputtering, grain-grain collisions, and accretion (Sect.~\ref{subsec:methods_dust}). Using this workflow, we are able to estimate the dust destruction efficiency of SNR forward shocks in different ISM environments.

\subsection{Turbulence driving}\label{subsec:methods_turbulence}
The 3D turbulence simulations were carried out purely hydrodynamically by using the turbulence driving within \texttt{Arepo} \citep{Mocz2017, Mocz2018}, which makes use of the Ornstein-Uhlenbeck process. We started the simulations with a homogeneous medium in which we drove the turbulence until it was in a statistically stationary state, and thus until the density and velocity PDF, and the power spectra converged. This method of driving let us generate a turbulence that closely resembles the observed ISM, as a visual comparison in Fig.~\ref{fig:turbulence_comparison} shows. There, the dust column density of one of our driven turbulence snapshots is qualitatively compared to a dust map of the Polaris flare from Herschel/SPIRE observations \citep{Miville-Deschenes2010}\footnote{\label{footnote1}We downloaded and visualized the level 2.5 data products from the Herschel Science Archive.} to show that the morphological features and scales are similar.

\begin{figure}[hbtp!]
    \centering
      \includegraphics[width=\linewidth]{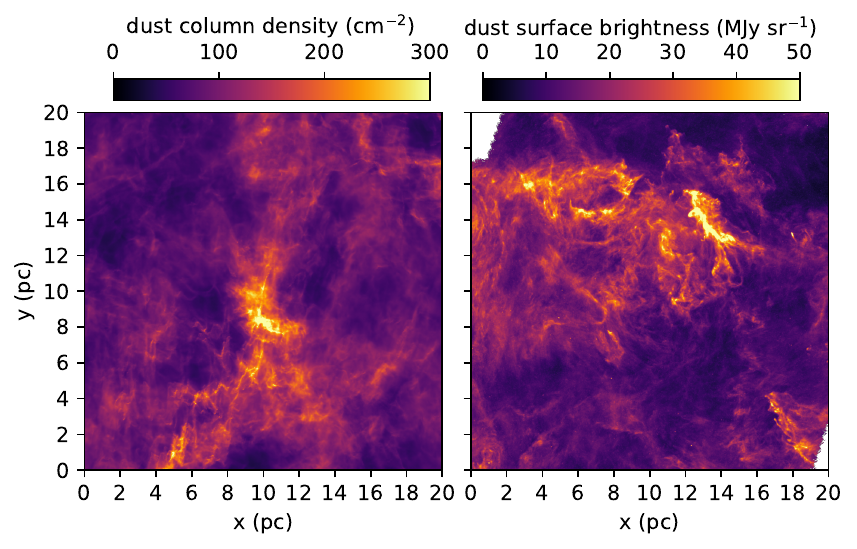}
    \caption{Structural comparison of the dust distribution in a turbulence snapshot of our intermediate-density (\unit[10]{cm$^{-3}$}) supersonic turbulence ($\mathcal{M}=3$) simulation (left) and an observed SPIRE \unit[250]{$\muup$m} dust map from Herschel observations of the Polaris Flare \citep[right, originally observed by][]{Miville-Deschenes2010}$^{\ref{footnote1}}$, which has a Mach number of $\mathcal{M}=2$--10 \citep{Beattie2019}. The physical length scale of \unit[$20\times20$]{pc} of the observed ISM is based on the distance of \unit[350]{pc} to the Polaris Flare  \citep{Schlafly2014, Panopoulou2022}.}
    \label{fig:turbulence_comparison}
\end{figure}

In the turbulence simulations the mass of every Voronoi cell was kept approximately the same, which induced refinement but did not change the total number of cells. We did not include additional refinement criteria for the turbulence driving. The box itself was \unit[20 or 40]{pc} large in each coordinate direction, depending on the used ISM density and turbulence strength. These box sizes were large enough so that the SNR could evolve for up to \unit[10]{kyr} without reaching the boundaries in every simulation. The boundaries were chosen to be periodic to ensure that no dust escapes from the box in post-processing. To be able to resolve filaments in the ISM that are observed to have a width of \unit[$\sim$0.1]{pc} \citep{Arzoumanian2011}, we used $512^3$ cells in total. Thus, the resolution in filaments was up to \unit[0.008]{pc} (\unit[0.016]{pc}) whereas in voids it was \unit[0.15]{pc} (\unit[0.3]{pc}) for the \unit[20]{pc} (\unit[40]{pc}) box.

The studied parameter sets include different turbulence strengths, defined by the turbulent Mach number $\mathcal{M}=\sigma_v / c$ that is dependent on the velocity dispersion $\sigma_v$ and the sound speed $c$ \citep{Beattie2019}. This Mach number determines the statistical distribution of the density and velocity, as can be seen in Fig.~\ref{fig:PDF_rho} with the density PDF. Lower turbulent Mach numbers only show slight density variations (e.g., $\mathcal{M}=0.3$, blue line), whereas higher turbulent Mach numbers result in dense filaments and empty voids (e.g., $\mathcal{M}=3$, pink line). Because the turbulent Mach number of the interstellar and circumstellar medium around stars can differ significantly depending on the exact location of the star, we took different Mach numbers into account -- from subsonic ($\mathcal{M}=0.3$) to transonic ($\mathcal{M}=1$) and supersonic ($\mathcal{M}=3$), and compared them to the homogeneous case ($\mathcal{M}=0$). This Mach number range reflects typical conditions in the ISM, with observed values ranging from subsonic (a few tenths) to supersonic \citep[e.g.,][$\mathcal{M}=0.17$--8]{Heyer2006,Ginsburg2013,Schneider2013, Marchal2021}.

\begin{figure}[hbtp!]
    \centering
    \includegraphics[width=.8\linewidth]{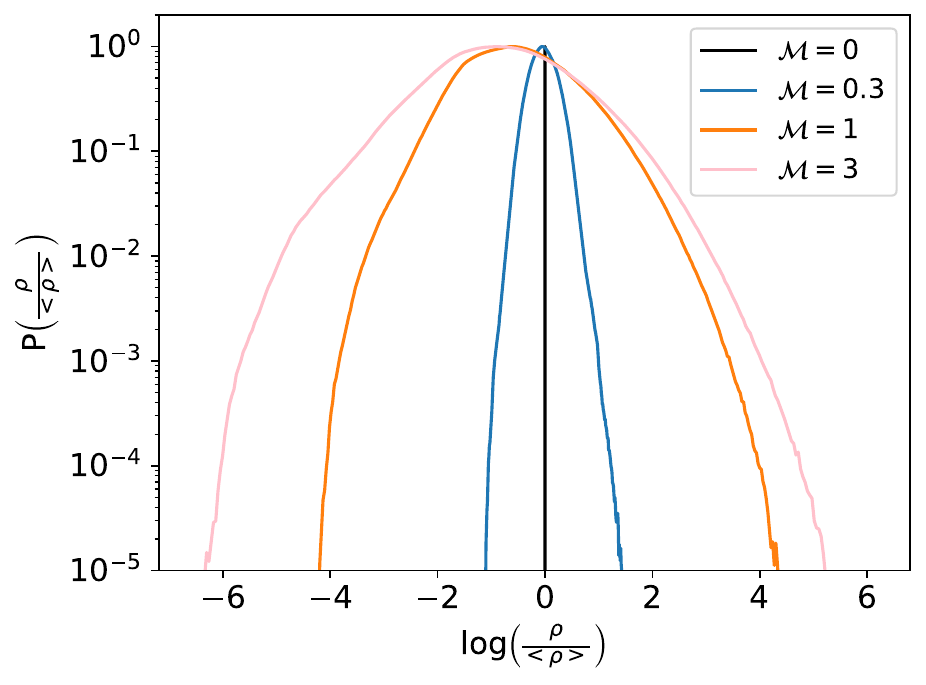}
    \caption{Density PDFs of the different turbulent Mach numbers $\mathcal{M}$ in the converged low-density (\unit[1]{cm$^{-3}$}) turbulence simulations. With a higher turbulent Mach number, cell densities ($\rho$) can have a larger deviation from the average density <$\rho$>.}
    \label{fig:PDF_rho}
\end{figure}

We also included different ISM densities since SNe can occur in media of various densities. Stars can remain in their birth molecular cloud throughout their lifetime, where they can explode close to dense filaments or in a medium that is emptied by previous SNRs, or can drift away from the cloud to lower density media \citep{MaykerChen2023, Sarbadhicary2023}. Thus, we chose to study different average ISM number densities of $n=\unit[1,10\text{ and }100]{cm^{-3}}$ in addition to the set of studied turbulent Mach numbers to comprehensively cover various realistic SNR surroundings. The column density of such a turbulence simulation box can be seen in Fig.~\ref{fig:turbulence_comparison} (left). Furthermore, slices of all simulations that were carried out in this study are visualized in Appendix~\ref{sec:appendix_SNRevo}, in Figs.~\ref{fig:gas_pics_ISM1}, \ref{fig:gas_pics_ISM10}, and \ref{fig:gas_pics_ISM100} for the low-, intermediate-, and high-density environments, respectively.

The initial ISM temperature in these simulations was chosen to be $\unit[10^4]{K}$. In the interest of maintaining a smaller computational requirement, we did not include magnetic fields, cooling or self-gravity in the turbulence driving. We did include the first two a posteriori in the SN explosion simulations, as laid out later in this section.

Since a magnetic field can play an important role in dust transport and destruction, we added it to the converged turbulent system with the same direction as the velocity vectors and a strength proportional to the number density $n$ with $B=bn^{1/2}$ \citep{Crutcher1999}, including the constant factor $b$. This factor was estimated to $b=\unit[0.93\pm0.87$]{$\muup \text{G}$\,cm$^{3/2}$} by averaging several observations of the magnetic field strength in dense filaments \citep{Curran2007, Chapman2011, Pattle2018, Coude2019, Liu2019, Wang2019, Arzoumanian2021, Eswaraiah2021, Konyves2021, Lyo2021, Ngoc2021, Pattle2021, Ching2022, Kwon2022, Karoly2023, Ward-Thompson2023, Wang2024}. We tested the impact of the direction of the magnetic field by comparing to a simulation with a 90$^\circ$ rotated magnetic field. The dust destruction results only differed by less than 0.1\% so that the magnetic field strength seems to have a larger impact than its direction for our models.

\subsection{SN explosion and SNR evolution}\label{subsec:methods_SNR}
The simulation output of Sect.~\ref{subsec:methods_turbulence} was used as the starting point for our MHD SNR simulations. In these simulations, we set off an SN with an explosion energy of \unit[10$^{51}$]{erg} using the SN feedback method of \texttt{Arepo} that injects thermal energy into 40 central neighboring Voronoi cells, corresponding to a radius of \unit[0.08--0.21]{pc} in our simulations. The SNR evolution was followed for \unit[10]{kyr}. We show the final snapshots in Fig.~\ref{fig:all_sims_10kyr} (see also Appendix~\ref{sec:appendix_SNRevo} for the full evolution of the simulations, including movies). In addition to the mass refinement of \texttt{Arepo}, we implemented refinement with respect to the pressure gradient to follow the SNR forward shock in greater detail, since it is the key region where dust destruction happens. Cell splitting and merging were enabled, so that the total number of cells almost doubled throughout the simulation because the SNR grew. Thus, the average distance between neighboring cell centers at the end of the simulations was of the order of \unit[0.030 and 0.060]{pc} for the \unit[20 and 40]{pc} box size, respectively.

\begin{figure*}[hbtp!]
    \centering
    \includegraphics[width=.77\linewidth]{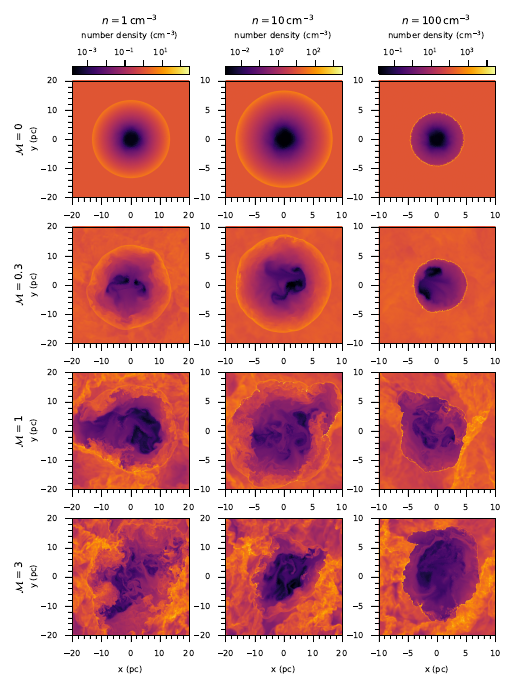}
    \caption{Number densities of the $z=0$ slice of all simulations carried out in this paper at \unit[10]{kyr}. This includes different average ISM number densities ($n$) and turbulent Mach numbers ($\mathcal{M}$). The colorbar scales are consistent in each of the average ISM number density cases but the box sizes can vary between \unit[20 or 40]{pc}.}
    \label{fig:all_sims_10kyr}
\end{figure*}

In the SNR simulations, radiative cooling is considered. We assumed collisional ionization equilibrium and electron-ion equipartition as we do not follow electron and ion temperatures separately, or ionization states in detail. The used cooling implementation in the code depends on whether the gas has been shocked, which was determined by its temperature. Unshocked ISM gas had an initial temperature of \unit[10$^4$]{K} and cooled according to detailed molecular and atomic cooling functions, assuming solar metallicity and a simplified treatment of gas-phase molecular chemistry \citep{Glover2012, 2016Bertram}. 
High-density filaments cool down efficiently before they are shocked, whereas voids remain at higher temperatures, due to their low density. We identified shocked gas in cells where the temperature exceeded \unit[10$^4$]{K}. We assumed that this gas mixed with SN ejecta, with the latter dominating the cooling process. Since our simulations did not include explicitly injected ejecta, the shocked ISM effectively took on the role of the ejecta. To emulate the conditions in the ejecta of Cassiopeia A (Cas A), which we used as a reference, we adopted oxygen-dominated cooling based on the \texttt{CHIANTI} cooling function \citep{Dere1997, DelZanna2015}, following the approach of \citet{Kirchschlager2019}. This method was chosen because applying the ISM cooling prescription to the ejecta resulted in a faster expansion of the SNR, due to insufficient cooling, leading to a divergence from the observed evolution of SNRs similar to Cas A. This approximation is considered reliable only during the early stages of the SN, when the shocked mass remains comparable to the expected ejecta mass. Because the swept-up ISM material is heated beyond the ISM cooling threshold, it was cooled according to the ejecta prescription, resulting in a slight overestimation of cooling at the shock front, particularly at late times in the high-density simulations.

We did not consider thermal conduction. Its timescale can be calculated by \citep[e.g.,][]{Orlando2005}
\begin{align}
    t_\text{cond} = \frac{7}{2(\gamma-1)} \frac{nk_Bl^2}{\kappa(T)},
\end{align}
where $l$ is the length scale of interest and $\kappa(T)=\unit[5.6\times10^{-7}T^{-\frac{5}{2}}]{erg\,s^{-1}\,K^{-1}\,cm^{-1}}$ is the thermal conductivity. The shortest timescale of \unit[$\sim$13]{kyr} is reached at the contact of the thin shell and the SNR interior of the high ISM density simulations. However, the forward shock is expanding on shorter timescales in our simulations so that thermal conduction can be neglected. Although at very early times of a few hundred  years, the thermal conduction timescale between the forward shock and the ISM is on the order of only a few thousand years, the ISM is heated more rapidly by the expansion of the forward shock than by conduction at this early stage.

Having set all initial parameters, the SN was set off in the driven turbulent ISM box. Because stars usually excavate their surroundings with stellar winds and radiative feedback, most of them explode in a low-density cavity \citep[e.g.,][]{Lamers1999, Rogers2013, Fichtner2024}. Modeling the stellar feedback mechanisms in detail is beyond the scope of this paper; however, we mimicked the impact of pre-SN feedback by always placing the SN in a void of the turbulence, also ensuring that all simulations are easily comparable. The void was manually chosen by shifting the 3D turbulence box to a position with a large void in the center. Therefore, the forward shock encountered filaments only after several hundred to a few thousand years of evolution for the higher Mach number simulations as is also the case for SNe that explode in a wind-driven bubble. The voids in our strong turbulence simulations were roughly \unit[10--20]{pc} large, which is comparable to the observed and predicted wind-driven cavity sizes of \unit[10--100]{pc} of massive stars, strongly dependent on the stellar mass and ISM density \citep{McCray1983, Georgy2013}. If the SNe were to explode closer to dense filamentary structures instead of voids, the filaments would be encountered at higher velocity, resulting in more efficient dust destruction in the early SNR stages. This is discussed in Appendix~\ref{sec:appendix_SN_at_filament}.

After evolving the SNR for \unit[10]{kyr} in MHD simulations, we studied the dust dynamics and destruction in post-processing simulations using \texttt{Paperboats} since \texttt{Arepo} does not include detailed dust processing. These dust simulations are run from snapshots of the SNR simulation, meaning that their time steps are larger than the hydrodynamic time steps. Since the forward shock is faster in the beginning and slows down significantly (from $\unit[v>10^4]{km\,s^{-1}}$ to $\unit[v<400]{km\,s^{-1}}$), the first few hundred years of evolution must have a higher temporal resolution than the last few hundred years. Thus, we took 512 snapshots of the MHD simulations at logarithmically spaced time steps to feed the post-processing simulations, resulting in time steps of \unit[0.15--120]{yr} for the dust simulations. 

\subsection{Dust dynamics}\label{subsec:methods_dust}
We post-processed the 512 snapshots from each MHD simulation with the \texttt{Paperboats} code \citep{Kirchschlager2019} in order to determine the dust dynamics. This code was developed to study the dust destruction of the reverse shock in dense ejecta clumps \citep{Kirchschlager2019, Kirchschlager2020, Kirchschlager2023, Kirchschlager2024b} and was later also used for forward shock dust destruction \citep{Kirchschlager2022, Kirchschlager2024a}. It simulates the dust by introducing a certain dust mass according to an initially fixed gas-to-dust mass ratio parameter that we adopted to be 100, which is close to the value assumed in dust models characterizing the diffuse ISM in the Milky Way \citep[e.g.,][]{Draine2007, Compiegne2011, Jones2013}. Based on the amount of dust, its properties, and the fluid conditions from the MHD simulations, \texttt{Paperboats} can determine how the dust will be affected by the gas and other dust grains via gas and plasma drag, Lorentz acceleration, thermal and kinetic sputtering, effects of grain-grain collisions (fragmentation and vaporization), and accretion in the following MHD snapshots. The rotational disruption of dust grains by mechanical torques (METD) or radiation \citep[RATD,][]{Hoang2020} is not included in \texttt{Paperboats}. These effects were recently evaluated in the context of SNR dust destruction \citep{Martinez-Gonzales2025}, and were found to only have a minor impact on the overall dust destruction. However, this could change for our model setup and will be part of analysis in the near future.

Initially, the dust was given the same velocity as the gas. As such, dust in the central cells would not be given an initial velocity so that it would not move out of the center. Therefore, we did not start with the first snapshot of the MHD simulation but from snapshot 10, which relates to only a few years after the explosion. The dust was then added with the dust grain-size distribution of Mathis, Rumpl and Nordsieck \citep[MRN;][]{Mathis1977} from 5 to \unit[250]{nm}, and thus a power-law distribution $a^{-\gamma}$ with $\gamma=3.5$. To account for grain destruction and accretion, the grain sizes in our simulations ranged from 0.6 to \unit[350]{nm}, which were separated into 20 logarithmically spaced bins. In total, we performed 4 dust destruction simulations for each SNR simulation, considering only sputtering, or sputtering and grain-grain collisions to assess the impact of grain-grain collisions, and either silicate or carbonaceous dust to study how the dust destruction efficiency varies for different grain species. More detailed information about the code implementation of \texttt{Paperboats} can be found in \citet{Kirchschlager2019, Kirchschlager2020, Kirchschlager2023}.

The gas-dust interactions simulated by \texttt{Paperboats} assume that the ISM gas consists of a single atomic species. We modeled the gas as being dominated by hydrogen atoms and therefore used a mean molecular weight of $\mu = 1$ for our dust simulations. As a result, hydrogen atoms must also dominate in the region where ISM gas and ejecta gas mix. This approximation becomes increasingly accurate as the SNR evolves since the mass of ISM gas in the SNR interior increases over time. In our simulations, the remnant encountered at least \unit[150]{M$_\odot$} of ISM material during \unit[10]{kyr}, which is significantly more than typical SNR ejecta masses (e.g., \unit[$\sim$4]{M$_\odot$} for Cas A). Thus, hydrogen dominated the total mass in the SNR interior for most of the evolution, although the cooling was still dominated by oxygen. For consistency, we used this simplified molecular weight for our density calculations in both the MHD and dust simulations. However, solar abundances were generally assumed for the rest of the MHD calculations.

Since \texttt{Paperboats} does not support a moving mesh yet, we converted the Voronoi grid of \texttt{Arepo} to a Cartesian grid that could be used by \texttt{Paperboats}. The Cartesian grid had 900 cells in each coordinate direction, almost twice as high as the initial average distance of cell centers in the MHD simulations. The dust simulation resolution was found to be optimal for our setup and to not induce significant errors during the simulations, as also discussed in Appendix~\ref{sec:appendix_4500cells}. During the conversion from Voronoi to Cartesian grid, the total mass of the system changed due to the rearrangement of the cells. However, our chosen resolution was high enough to only change the mass by 0.2\%, which had a minor impact on the dust destruction. 

Furthermore, we only studied the central 2D layer because a full 3D dust calculation would have been numerically too expensive. The final amount of Cartesian grid cells was therefore $900\times 900\times 1$ cells in $x$-, $y$-, and $z$-direction. To ensure that the investigated layer was representative of the whole 3D SNR, we performed the dust simulations on three different perpendicular planes. The dust destruction of the forward shock in these planes can differ by several per cent as shown in Appendix~\ref{sec:appendix_diff_slices}, depending on the different detailed density distribution in the slice. Thus, dust destruction values in this paper always refer to the average of these three planes, unless otherwise stated.

To evaluate the dust destruction efficiency of only the encountered ISM dust instead of the dust in the full simulation box, we had to trace the region which was shocked by the SNR in all snapshots. Since the SNR is still hot and fast after \unit[10]{kyr}, it was possible to follow it by using a temperature or velocity gradient threshold, which were implemented into \texttt{Paperboats}. The temperature threshold worked best for the subsonic and homogeneous cases and traced all cells which are or were above the temperature of $\unit[1.1\times10^4]{K}$ in any of the MHD snapshots. This threshold was chosen to correctly follow the SNR in all time steps, even in the first few where the ISM is at $\unit[10^4]{K}$ but cools down rapidly. For the trans- and supersonic simulations, voids did not cool quickly but can be compressed and thus heated by the slight diffusion of the filaments. Therefore, we increased the temperature threshold to $\unit[3\times10^4]{K}$ for the trans- and supersonic simulations with a low ISM density. For the intermediate- and high-density cases, however, it is possible that, at a given snapshot, shocked gas cells cool down to temperatures lower than the ones retained in the unshocked gas in voids. Hence, we used a $v\Delta v_i$ threshold that traced all cells that are or were above it at any given time step for the trans- and supersonic simulations of these cases. This $\Delta v_i$ threshold is the velocity difference of neighboring cells in the coordinate direction $i\in \{x,y,z\}$. Because this threshold is resolution dependent and the forward shock evolves differently depending on the ISM density and Mach number, we adjusted the $v\Delta v_i$ threshold for each of the four simulations for which it was used. This was done by carefully fine-tuning the threshold until no ISM gas cells but all SNR forward shock cells were traced. Since we tracked all cells that are or once were above this threshold, we finally followed the whole SNR interior over time. We manually checked that the SNR evolution was followed correctly for all of our snapshots. This meant that we could follow the dust destruction and grain-size distributions of the SNR shocked region more easily, noting that mixed cells at the shock front may be traced too early or too late. However, cells at the boundary were only a small fraction of all traced SNR cells so that the impact on the total dust destruction rate is negligible.

\section{Results}\label{sec:results}
In this section we discuss the impact of the different initial ISM densities and turbulent Mach numbers on the SNR evolution, and analyze how they influence the dust destruction efficiency, considering different dust species and destruction mechanisms. The gas density distribution of all SNR simulations at \unit[10]{kyr} is shown in Fig.~\ref{fig:all_sims_10kyr} (see Appendix~\ref{sec:appendix_SNRevo} for the full evolution, including movies). Depending on the considered parameter set, the total dust destruction lies between 27--92\%, corresponding to \unit[0.85--11.0]{M$_\odot$} of destroyed dust in the first \unit[10]{kyr} of SNR evolution. The given destroyed dust masses are not the dust masses in a 2D plane but an extrapolation of the 2D destruction rates multiplied with the 3D encountered dust mass. The dust destruction results, averaged over the three investigated orthogonal planes, are compared in Table~\ref{tab:all_sims} and the results of the $z=0$ plane dust simulations are visualized in Fig.~\ref{fig:destr_results} where the rows show the different Mach numbers ($\mathcal{M}=0,0.3,1,3$) and the columns show the average ISM densities ($n=\unit[1,10,100]{cm^{-3}}$).

\begin{table*}[hbtp!]
    \centering
    \caption{\label{tab:all_sims}Overview of the results after \unit[10]{kyr} for all dust simulations carried out in this paper.}
    \begin{tabular}{cccccccc}
        $n$ (cm$^{-3}$) & $\mathcal{M}$ & Si destr. (\%) & Si destr. (M$_\odot$) & C destr. (\%) & C destr. (M$_\odot$) & $v_\text{shock}^\text{end}$ $ (\text{km\,s}^{-1})$ \\\hline
        1   & 0 & 56.8 & 1.46  & 35.5 & 0.91  & 388$\pm$3\\
        1   & 0.3& 53.0 $\pm$ 1.8 & 1.37 $\pm$ 0.05 & 33.0 $\pm$ 1.1 & 0.85 $\pm$ 0.03 & 348$\pm$3\\
        1   & 1 & 45.6 $\pm$ 2.7 & 1.53 $\pm$ 0.09 & 29.0 $\pm$ 1.5 & 0.97 $\pm$ 0.05 & 247$\pm$18\\ 
        1   & 3 & 42.7 $\pm$ 3.6 & 1.49 $\pm$ 0.13 & 27.3 $\pm$ 2.7 & 0.95 $\pm$ 0.09 & 251$\pm$26\\\\

        10  & 0 & 91.7 & 5.63 & 76.9 & 4.72 & 226$\pm$1\\ 
        10  & 0.3&90.8 $\pm$ 0.1 & 5.62 $\pm$ 0.01 & 76.3 $\pm$ 0.1 & 4.72 $\pm$ 0.01 & 210$\pm$6\\ 
        10  & 1 & 78.3 $\pm$ 0.5 & 5.25 $\pm$ 0.03 & 63.0 $\pm$ 1.6 & 4.23 $\pm$ 0.11 & 198$\pm$16\\ 
        10  & 3 & 57.6 $\pm$ 2.1 & 4.43 $\pm$ 0.16 & 45.5 $\pm$ 3.9 & 3.50 $\pm$ 0.3 & 226$\pm$21\\\\ 
        
        100 & 0 &86.9 & 9.23 & 87.2 & 9.26 & 127$\pm$4\\ 
        100 & 0.3&86.9 $\pm$ 0.1 & 9.13 $\pm$ 0.02 & 87.0 $\pm$ 0.1 & 9.14 $\pm$ 0.01 & 128$\pm$3\\
        100 & 1 & 81.0 $\pm$ 0.6 & 9.41 $\pm$ 0.06 & 79.2 $\pm$ 0.6 & 9.20 $\pm$ 0.06 & 125$\pm$5\\ 
        100 & 3 & 75.9 $\pm$ 0.9 & 11.0 $\pm$ 0.14 & 73.0 $\pm$ 1.5 & 10.58 $\pm$ 0.22 & 88$\pm$1\\

        \\ 
        \hline

        1   & 0 & 55.5 & 1.42 & 33.0 & 0.85 & 388$\pm$3\\
        1   & 0.3&51.7 $\pm$ 1.8 & 1.34 $\pm$ 0.05 & 30.4 $\pm$ 1.1 & 0.79 $\pm$ 0.03 & 348$\pm$3\\ 
        1   & 1 & 44.8 $\pm$ 2.7 & 1.5 $\pm$ 0.09 & 27.1 $\pm$ 1.5 & 0.91 $\pm$ 0.05 & 247$\pm$18\\ 
        1   & 3 & 41.8 $\pm$ 3.7 & 1.46 $\pm$ 0.13 & 25.3 $\pm$ 2.6 & 0.88 $\pm$ 0.09 & 251$\pm$26\\\\ 

        10  & 0 & 91.0 & 5.58 & 74.2 & 4.55 & 226$\pm$1\\ 
        10  & 0.3&90.1 $\pm$ 0.1 & 5.57 $\pm$ 0.01 & 73.7 $\pm$ 0.1 & 4.56 $\pm$ 0.01 & 210$\pm$6\\ 
        10  & 1 & 75.4 $\pm$ 0.9 & 5.05 $\pm$ 0.06 & 55.8 $\pm$ 0.4 & 3.74 $\pm$ 0.03 & 198$\pm$16\\ 
        10  & 3 & 50.2 $\pm$ 2.2 & 3.86 $\pm$ 0.17 & 33.1 $\pm$ 1.2 & 2.55 $\pm$ 0.09 & 226$\pm$21\\\\ 
        
        100 & 0 &69.8 & 7.41 & 63.1 & 6.7 & 127$\pm$4\\ 
        100 & 0.3&70.6 $\pm$ 0.5 & 7.43 $\pm$ 0.06 & 64.2 $\pm$ 0.6 & 6.75 $\pm$ 0.06 & 128$\pm$3\\ 
        100 & 1 & 65.2 $\pm$ 3.1 & 7.57 $\pm$ 0.36 & 55.9 $\pm$ 3.2 & 6.49 $\pm$ 0.37 & 125$\pm$5\\ 
        100 & 3 & 49.6 $\pm$ 0.8 & 7.18 $\pm$ 0.11 & 36.6 $\pm$ 0.7 & 5.30 $\pm$ 0.10 & 88$\pm$1 
    \end{tabular}
    \tablefoot{The simulations vary in their considered average ISM density $n$, turbulent Mach number $\mathcal{M}$, and dust species. The values in the top 12 lines belong to the simulations that consider grain-grain collisions and sputtering, while the bottom 12 lines belong to the sputtering-only simulations. The values are averaged over three orthogonal 2D planes. Their standard deviation from the mean was used as the error of the mean. The average end shock velocity $v_\text{shock}^\text{end}$ was estimated by locating the SNR forward shock from a momentum threshold. The uncertainty in this value is due to the use of different thresholds, all of which trace the shock reasonably well.}
\end{table*}

\begin{figure*}[hbpt!]
  \centering
  \includegraphics[width=.9\linewidth]{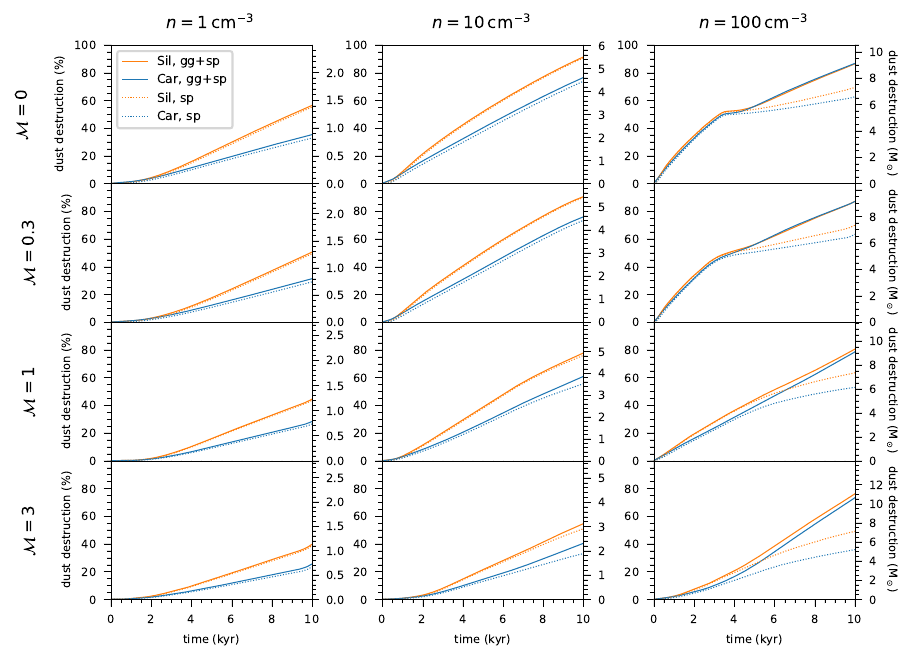}
  \caption{Dust destruction efficiencies (left $y$-axis) and masses (right $y$-axis) in the $z=0$ slice. This destruction efficiency refers to the ratio of the destroyed dust mass at each time step to the total encountered dust mass at the last considered snapshot. Different Mach numbers of $\mathcal{M}=0,0.3,1,\text{ and }3$, and average ISM densities of $\unit[n=1,10,\text{ and }100]{cm^{-3}}$ are shown. For each Mach number and ISM density parameter set, four different dust simulations were carried out, including either silicate (Sil) or carbonaceous (Car) dust, and either grain-grain collisions and sputtering (gg+sp) or only sputtering (sp).}
  \label{fig:destr_results}
\end{figure*}

\subsection{Low-density ISM}\label{subsubsec:SNR_low_dens}
In a low-density environment ($n=\unit[1]{cm^{-3}}$, left column of Fig.~\ref{fig:all_sims_10kyr}), the forward shock is not efficiently slowed down by ISM gas (\unit[247--388]{km\,s$^{-1}$} at \unit[10]{kyr}, Table~\ref{tab:all_sims}) and can thus expand to large sizes, reaching up to \unit[24--40]{pc}, depending on the studied turbulent Mach number. 
Due to the low amount of shocked gas in this ISM and the fixed initial gas-to-dust mass ratio, small amounts of dust are encountered, resulting in low destroyed dust masses of \unit[1.4-1.5]{M$_\odot$} for silicate dust, and \unit[0.85--0.97]{M$_\odot$} for carbonaceous dust after \unit[10]{kyr} in our simulations. The destruction efficiencies are also rather low with 43--57\% for silicate and 27--36\% for carbonaceous dust (also shown in Table~\ref{tab:all_sims} and Fig.~\ref{fig:destr_results}), which may be unexpected given that the shock front is fast and hot throughout the whole simulation, implying efficient thermal and kinetic sputtering. However, sputtering of dust grains can only occur when they collide with the hot and fast ions of the shock front. Since these ions have a rather low number density and move through a low-density ISM, which has an even lower dust grain number density of \unit[0.01]{cm$^{-3}$} on average, the energetic ions of the forward shock do not collide often with dust grains. Thus, they cannot transfer their energy to the ISM grains (i.e., sputter them) efficiently. Due to the low grain number density, grain-grain collisions rarely occur, leading to differences of maximum 2.6\% compared to the sputtering-only simulations. Therefore, the processed grain-size distributions with and without grain-grain collisions are similar for grains larger than \unit[5]{nm}, showing a survival of most grains above \unit[$\sim$40]{nm} and significant destruction fractions for smaller grains. Many grains smaller than \unit[5]{nm} are generated, increasing even more with included grain-grain collisions that produce many small fragments (see Appendix~\ref{sec:appendix-GSDs}).

The filaments in the high Mach, \unit[$n=1$]{cm$^{-3}$} simulations are much denser than the average ISM, but their density is mostly below \unit[100]{cm$^{-3}$}, which is less than the average density in our high ISM density cases. Thus, they are not dense enough to stop the forward shock but get penetrated by it. This can be seen in Fig.~\ref{fig:zoom_fingers}, which shows the interaction of the forward shock with a filament in the $n=\unit[1]{cm^{-3}}$ and $\mathcal{M}=1$ case. The filament does not preserve its integrity but is processed by the forward shock. Fingers form from the destroyed filament into the ejecta, not only at the shown filament, but also at other forward shock positions (see Fig.~\ref{fig:all_sims_10kyr}).

\begin{figure}[hbtp!]
    \centering
    \includegraphics[width=\linewidth]{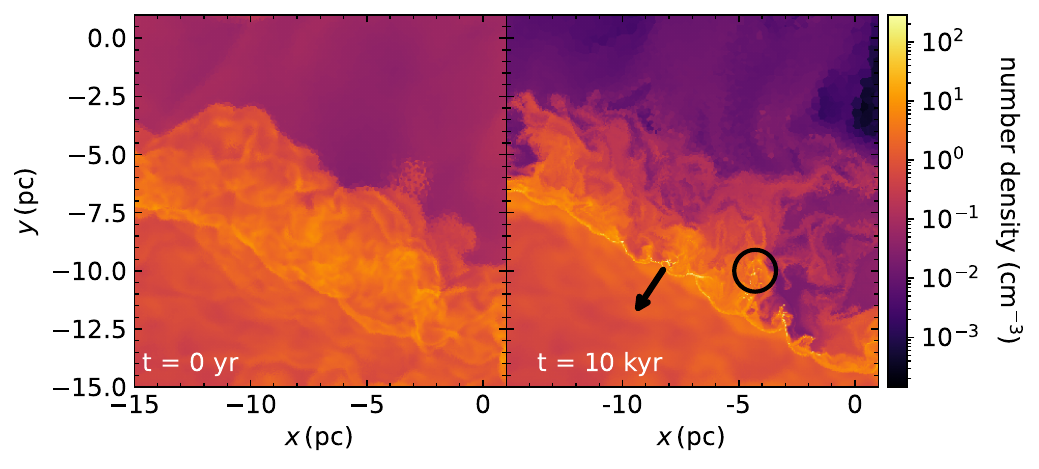}
    \caption{\unit[$16\times16$]{pc} zoomed-in image of a filament of the $n=\unit[1]{cm^{-3}}$ and $\mathcal{M}=1$ simulation at $t=\unit[0]{yr}$ and $t=\unit[10]{kyr}$. The filament cannot withstand the shock efficiently, but is destroyed leaving behind finger-like structures that arise from the forward shock front into the SNR interior. Furthermore, some high-density cloudlets are formed (circled in black). The arrow indicates the approximate direction of the forward shock.}
    \label{fig:zoom_fingers}
\end{figure}

A few high-density cloudlet structures can be seen that are a few pixels wide and have higher densities than the initial filaments ($n>\unit[100]{cm^{-3}}$, e.g., the ones circled in Fig.~\ref{fig:zoom_fingers}). These cloudlets result from compressed material of the processed filament and can shield some dust from destruction as they take longer to be disrupted than the lower density surroundings.

Although the filaments do not withstand the energetic forward shock but get processed to lower density fingers and eventually destroyed, they are able to shield some dust in comparison to the homogeneous case in the most destructive, first \unit[10]{kyr} because it takes longer for them to be penetrated and processed by the shock than in the homogeneous case. Roughly 14\% of silicates and 8\% of carbonaceous dust grains are shielded by the filaments in the supersonic turbulent Mach number case, decreasing with turbulence strength to 4\% of silicate and 3\% of carbonaceous dust for a subsonic Mach number as can be seen in Fig.~\ref{fig:destr_results_same_density_different_M}. This supports the conclusions of previous papers with a longer evolution time that find that filaments or inhomogeneities in general are able to shield dust from destruction\break \citep{Hu2019, Kirchschlager2024a, Dedikov2025}. However, the destroyed dust mass remains on the same order of magnitude for all Mach number simulations since the forward shock encounters more dust in the high Mach number cases.

\begin{figure*}[hbpt!]
  \centering
  \includegraphics[width=.8\linewidth]{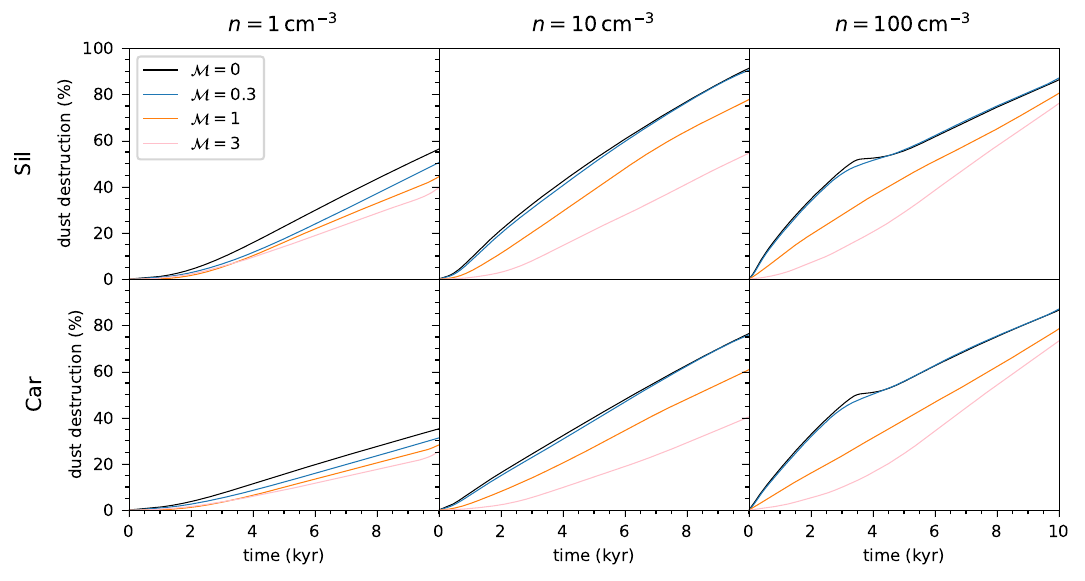}
  \caption{Dust destruction efficiencies in the $z=0$ slice in a direct comparison between the different Mach numbers ($\mathcal{M}$) used for each ISM density ($n$) case, considering either carbonaceous (Car) or silicate (Sil) grains.}
  \label{fig:destr_results_same_density_different_M}
\end{figure*}

This happens due to the turbulent ISM structures in the high Mach number cases that lead to an irregular expansion of the SNR. While it can rapidly move through the voids that have a very low density, the forward shock is slowed down more in higher density regions. Since filaments are spatially thin in comparison to the voids, the shock can evolve through lower density areas around them \citep[cf.][]{Korolev2015, Slavin2017, Dedikov2025}. Thus, after \unit[10]{kyr}, the forward shock undergoes greater expansion in the high Mach number ISM than in the homogeneous case and is also more efficiently slowed down on average across the entire shock, due to its low velocity in the filaments. 

The velocity of the forward shock, which can be estimated by averaging the gas velocity of the shock cells traced by a momentum threshold, is therefore still \unit[388]{km\,s$^{-1}$} in the homogeneous case, whereas it slows down to \unit[251]{km\,s$^{-1}$} in the supersonic turbulence case after \unit[10]{kyr}, as given in Table~\ref{tab:all_sims}. Since kinetic sputtering is directly dependent on the forward shock velocity, a comparison of the simulations at a similar end shock velocity instead of similar end time would also be a valuable comparison. This was done in Appendix~\ref{sec:appendix-same_FS_v}, where all simulations are compared at different evolution times but similar average shock velocities of \unit[387--390]{km\,s$^{-1}$}. For the low-density simulations, the spread in dust destruction efficiency between the different Mach numbers increases to 26\% for the silicate grains and 16\% for the carbonaceous grains; however, the high Mach number simulations show significantly less destroyed dust mass in this comparison, reflecting the trend that the filaments can shield the dust from destruction.

Regardless of the comparison criterion of the simulations (Table~\ref{tab:all_sims} or \ref{tab:all_sims_same_shock_vel}) and the studied parameter set, silicates are always at least 12\% more efficiently destroyed than carbonaceous grains in a low-density environment. This difference can generally be explained by the higher sputtering yields of silicates that stem from their higher bulk density.\footnote{The bulk density of silicate and carbonaceous grains is \unit[3.3]{g\,cm$^{-3}$} and \unit[2.2]{g\,cm$^{-3}$}, respectively \citep{Tielens1994, Jones1994}.} Thus, the silicate grains are also less efficiently accelerated by the forward shock, resulting in higher relative dust-to-gas velocities and thus more efficient kinetic sputtering at the shock front. However, other mechanisms are also affected. Due to their higher bulk density, silicates are less numerous than carbonaceous grains since we consider the same dust sizes and total dust mass for both species, resulting in fewer grain-grain collisions and a lower sputtering efficiency due to a lower total grain surface area. Furthermore, silicates have lower relative grain-to-grain velocities when they are accelerated by the forward shock so that fragmentation and vaporization in general are less efficient for silicates than for carbonaceous grains. Moreover, the size-dependency of sputtering is considered in \texttt{Paperboats} \citep[see][]{Kirchschlager2019}, which arises since smaller grain sizes are used in comparison to the experimental sputtering yields which considered a semi-infinite target \citep{Jurac1998, SerraDiaz-Cano2008}. This correction was updated in our study to consider hydrogen ions of the ISM to impinge on dust grains as projectiles instead of oxygen ions as used previously by \citet{Kirchschlager2022, Kirchschlager2024a}. It generally leads to fewer silicate grains but more small carbonaceous grains being destroyed, supporting a better survival of silicates. For the low-density simulations, however, the generally higher sputtering yields and more efficient kinetic sputtering during the acceleration of the grains seem to be more important than the higher sputtering efficiency due to a higher total grain surface area, the corrections of the size-dependent sputtering, and the higher fragmentation and vaporization efficiencies.

In total, more than \unit[1.3]{M$_\odot$} of silicate dust is destroyed in each low-density simulation in the first \unit[10]{kyr}, which is higher than the maximum amount of dust (\unit[$\sim$1]{M$_\odot$}) produced in SNRs as inferred from observations. However, this threshold is not reached in the case of carbonaceous grains yet: slightly less than \unit[1]{M$_\odot$} is destroyed in the first \unit[10]{kyr}. Although this seems to imply that SNRs are net dust destroyers for an ISM with predominantly silicate grains but can be net producers for a carbonaceous dust-dominated environment, the results of our simulations are only lower limits of the dust destruction, since we only investigate the first \unit[10]{kyr} of SNR evolution. At \unit[10]{kyr}, the dust destruction efficiency of all low-density simulations has not yet plateaued, meaning that the dust is still efficiently destroyed. Furthermore, the forward shock is still fast and hot at \unit[10]{kyr} (\unit[>240]{km\,s$^{-1}$}) so that it is reasonable to expect efficient dust destruction by kinetic and thermal sputtering beyond \unit[10]{kyr}. If the so far neglected reverse shock dust destruction would also be taken into account, it is probable that SNRs destroy more dust than they produce in this low-density environment, even if the ISM is dominated by carbonaceous dust.

We also note that, although we see locally reflected shocks that are generated at filaments and reverse through the SNR, a global reverse shock does not occur in any of our simulations. Since we do not include a mass injection with the SN explosion, which would incur a proper ejecta evolution, there is no ejecta mass that could compress the gas behind the forward shock. Thus, there is no pressure build-up leading to a global reverse shock, which could destroy ejecta dust. However, studying the reverse shock destruction of ejecta dust is beyond the scope of this paper.

\subsection{Intermediate-density ISM}
The SNR forward shock in the intermediate-density environment ($n=\unit[10]{cm^{-3}}$, central column of Fig.~\ref{fig:all_sims_10kyr}) is slowed down more efficiently than in the low-density environment, ending at \unit[198--226]{km\,s$^{-1}$} (Table~\ref{tab:all_sims}) and only reaching sizes of \unit[16--28]{pc}. In total, 58--92\% or \unit[4.4--5.6]{M$_\odot$} of silicate dust, and 46--77\% or \unit[3.5--4.7]{M$_\odot$} of carbonaceous dust is destroyed. Thus, carbonaceous grains are again more robust than silicate grains with similar differences of at least 12\% more destruction of silicates as in the low-density simulations. 

With more than \unit[3.5]{M$_\odot$} of destroyed dust for every intermediate-density simulation, significantly more dust is destroyed than in the low-density cases. The destruction efficiency is also higher since collisions of the hot and high-velocity ions with the ISM dust grains occur more often, resulting in more efficient thermal and kinetic sputtering. Grain-grain collisions are still not very efficient in the homogeneous and subsonic Mach number cases, leading to differences of $\sim$1\%. They make a bigger impact of 2.9--12\% for the more turbulent cases of this ISM density as they are typically more efficient in higher density media such as filaments. For all cases, similar to the low-density environment, the grain-size distribution with and without grain-grain collisions is similar for grains larger than \unit[5]{nm} but is altered for smaller grains (see Appendix~\ref{sec:appendix-GSDs}). In general, the final grain-size distribution is altered in a similar way as in the low-density cases, having more smaller grains destroyed than larger grains. However, as grain-grain collisions become effective in the high Mach number simulations, more and more large grains are destroyed.

Compared to the low-density ISM, the filaments in the intermediate-density transonic and supersonic cases have a higher number density, mostly on the order of \unit[1000]{cm$^{-3}$}. Thus, they can better withstand the forward shock that itself is generally more decelerated due to the higher average ISM density. When the filaments are encountered by the forward shock, they are not immediately destroyed, but partly swept up and dragged along with the shock, so that a thin and dense shell-like structure can form, which can cool down more efficiently (see Fig.~\ref{fig:all_sims_10kyr}). However, most of the filaments are still eventually destroyed, leaving behind finger-like structures that expand into the ejecta, similar to Fig.~\ref{fig:zoom_fingers}, and high-density cloudlets.

Because of the longer filament survival times and the lower penetration velocity, the dust is shielded much better in the dense regions of the higher Mach number cases, independent of the dust species, as shown in Fig.~\ref{fig:destr_results_same_density_different_M}. Thus, we can observe a large difference in dust destruction efficiencies between homogeneous and strongly turbulent ISM of up to 34\% for the silicate grain simulations and 31\% for the carbonaceous grain simulations, decreasing with Mach number to 0.9\% dust destruction difference with silicates and 0.6\% difference with carbonaceous grains. As a result, also less dust mass is destroyed by the SNR in the higher Mach number simulations, although it expands to larger sizes.

At similar end shock velocities of \unit[387--390]{km\,s$^{-1}$} instead of similar end times, this trend remains visible and accounts for up to 39\% and 34\% of dust destruction difference for the silicate and carbonaceous grains, respectively (Appendix~\ref{sec:appendix-same_FS_v}). The amount of destroyed dust mass also has a larger spread between the Mach number cases, and remains higher than in the low-density ISM; however, the dust destruction difference between intermediate- and low-density ISM cases decreases slightly. 

Although filaments are able to shield the dust from destruction, the SNR forward shock is strong enough to always destroy much more dust than \unit[1]{M$_\odot$} in the first \unit[10]{kyr}. Furthermore, the destruction efficiencies do not plateau yet, indicating that the dust is still rather efficiently destroyed at the end of our simulations. Thus, SNRs in this environment are net dust destroyers, even if the reverse shock would not destroy any produced dust and only the first \unit[10]{kyr} of evolution would be considered.

\subsection{High-density ISM}\label{subsubsec:SNR_high_dens}
With the highest studied average ISM density (\unit[100]{cm$^{-3}$}), we aim to replicate a molecular cloud environment with filament densities of the order of \unit[10$^4$]{cm$^{-3}$}. The SNR forward shock is strongly slowed down to \unit[88--128]{km\,s$^{-1}$} (Table~\ref{tab:all_sims}), thereby only reaching a maximum radius of \unit[10--17]{pc} in size after \unit[10]{kyr} (right column of Fig.~\ref{fig:all_sims_10kyr}). After \unit[10]{kyr}, 76--87\% or \unit[9.1--11.0]{M$_\odot$} of silicate dust, and 73--87\% or \unit[9.1--10.6]{M$_\odot$} of carbonaceous dust is destroyed (Fig~\ref{fig:destr_results}). 

In contrast to the intermediate- and low-density cases, the forward shock in the high-density case is compressed to a thin and dense shell from \unit[$\sim$3--4]{kyr} onward. This happens because larger amounts of gas are swept up than in the lower density cases, the shock is decelerated more to $v<\unit[130]{km\,s^{-1}}$, and the SNR interior cools down enough. The formation of the shell can be seen well in the high-density, homogeneous ISM simulation in Fig.~\ref{fig:all_sims_10kyr}, in which the SNR interior is empty whereas the shell has a very high density and starts to exhibit Vishniac thin-shell modes \citep{Vishniac1983, Vishniac1994}, which are inhibited by the irregular expansion of the forward shock in the higher Mach turbulence cases.

From the time the dense shell forms onward, grain-grain collisions become very efficient. They account for a maximum difference of 36\% in dust destruction at \unit[10]{kyr}, which is significantly more than in our low- and intermediate-density models. Due to the high density of the shell, the forward shock starts to cool down more efficiently, and thus slows down more so that the effectiveness of both thermal and kinetic sputtering decreases. At a certain critical velocity, which is dependent on the grain composition, the kinetic sputtering yield drops steeply, making it even less efficient \citep{Tielens1994, Nozawa2006}. Hence, after \unit[3--4]{kyr}, the sputtering-only simulations start to have a shallow slope in dust destruction (Fig.~\ref{fig:destr_results}). The simulations with all destruction mechanisms included continue to have high destruction efficiencies. Thus, with the onset of the shell, the significance of the destruction mechanisms alters since sputtering becomes inefficient whereas grain shattering gains in effectiveness.

Due to this change, the destruction efficiencies of carbonaceous and silicate grains in the high-density ISM are only 3\% different at most. For the weakly turbulent and homogeneous cases, carbonaceous grains are even slightly more efficiently destroyed than silicates. This partly results from the higher shattering efficiency of carbonaceous grains in comparison to silicate grains, as discussed in Sect.~\ref{subsubsec:SNR_low_dens}, and becomes evident from the larger differences of at least 6\% between the carbonaceous and silicate grain simulations with only sputtering considered, shown in Table~\ref{tab:all_sims}. However, as described in Sect.~\ref{subsubsec:SNR_low_dens}, several mechanisms play a role in destroying more carbonaceous than silicate grains in these particular simulations.

The grain-size distribution also changes more significantly in these high-density cases because larger dust grains are more efficiently destroyed by shattering than smaller grains. The processed grain-size distribution is therefore shifted to smaller grain sizes compared to the initial MRN distribution, as discussed in Appendix~\ref{sec:appendix-GSDs}.

Since sputtering becomes significantly less efficient with the formation of the shell, the trend of a higher dust destruction efficiency with higher ISM density also discontinues with the $\unit[n=100]{cm^{-3}}$ cases although more dust mass is destroyed because more dust is encountered. While the forward shock dust destruction efficiency in a high-density and low Mach number ISM is still higher than in the intermediate-density medium at \unit[3]{kyr}, it is roughly 4--5\% lower than in the intermediate-density cases at \unit[10]{kyr}. If only sputtering is considered, almost all other Mach number cases also decrease in destruction efficiency in the high-density ISM compared to the intermediate-density cases after \unit[10]{kyr}, by up to 21\%.

However, a comparison at the same end shock velocities as in Appendix~\ref{sec:appendix-same_FS_v} still shows the trend of more dust destruction with a higher average ISM density because the shell has not yet formed during the modeled period. Thus, a higher dust fraction and more dust mass is destroyed in the high-density ISM than in the intermediate-density cases. We note that this comparison has to be judged carefully since the SNR shock does not encounter many filamentary structures in the higher Mach number simulations yet, and thus, mostly evolves in a void with a density \unit[$n\lesssim$10]{cm$^{-3}$}, rather resembling an intermediate- or low-density case. Thus, the dust destruction in the higher Mach number simulations is also much lower than in the homogeneous ISM with large differences of up to 36\%.

Smaller differences are found after \unit[10]{kyr} of evolution time. The destruction efficiency from the homogeneous to the strongly turbulent ISM case is only reduced by 11\% for the silicate grains and 14\% for the carbonaceous grains, as seen in Fig.~\ref{fig:destr_results_same_density_different_M}. This is less than in the intermediate-density cases and similar to the low-density cases although the filaments are able to slow down the shock more efficiently and almost stop it at high densities. The forward shock shape at the end of our simulations is similar to the initial filament structures, as shown in Fig.~\ref{fig:zoom_filament_stopping}, and almost no finger-like shapes are formed but some dust-shielding cloudlets occur. However, due to the shell formation in the high-density cases and the subsequent shift of significant destruction mechanisms, the shielding ability of the filaments decreases. Since shattering is more efficient in higher density media, it shrinks the difference in dust destruction between the homogeneous and strongly turbulent cases by destroying dust in dense filaments more efficiently than in the homogeneous and weakly turbulent runs. Thus, at roughly \unit[4]{kyr}, the dust destruction efficiency difference between weak and strong turbulence is still on the same order of the intermediate-density cases with over 30\%, but decreases again after the shell forms.

\begin{figure}[hbtp!]
    \centering
    \includegraphics[width=\linewidth]{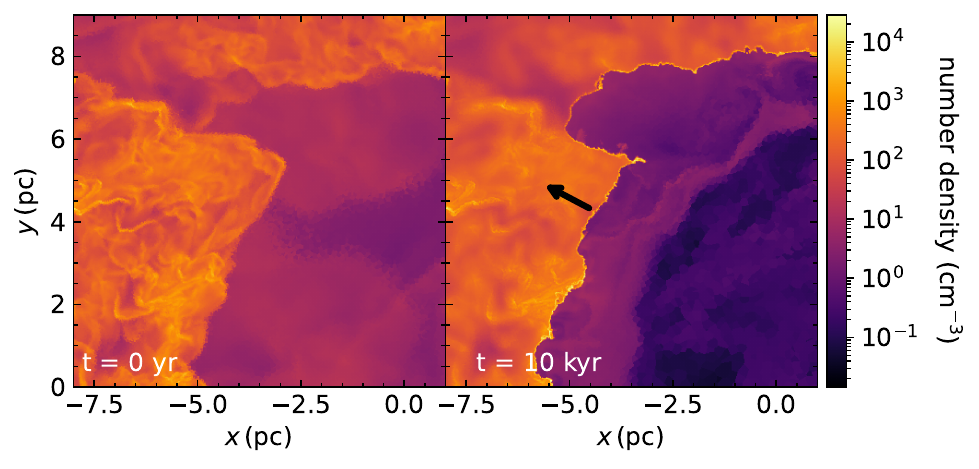}
    \caption{\unit[$9\times9$]{pc} zoomed-in image of a filament of the $\unit[n=100]{cm^{-3}}$ and $\mathcal{M}=3$ simulation at \unit[$t=0$]{yr} and \unit[$t=10$]{kyr}. The dense filament almost stops the shock and dictates where it will evolve. The initial filament shape looks nearly identical to the forward shock shape at the end. The arrow indicates the approximate direction of the forward shock.}
    \label{fig:zoom_filament_stopping}
\end{figure}

In total, SNRs always destroy more than \unit[9]{M$_\odot$} of dust in their first \unit[10]{kyr} of evolution in a high-density ISM, which is a magnitude more than what SNe have been observed to produce at most, making them significant net dust destroyers in this environment. An even higher dust destruction would be found if the considered ejecta cooling would be less efficient than our rather strong oxygen cooling so that the SNR would expand more and the shell would form later on, resulting in more dust being encountered and thus destroyed and a longer period of efficient sputtering.

\subsection{Summary}
To summarize, the SNR forward shock dust destruction depends on the properties of the surrounding medium (ISM density and turbulent Mach number), the dust grain species, and the considered destruction mechanisms. If only one of these parameters is varied, the dust destruction efficiency can change significantly. Generally, we find that the dust destruction efficiency decreases with lower ISM densities and higher turbulent Mach numbers. It is further reduced if carbonaceous instead of silicate dust is considered. Moreover, grain-grain collisions always generate many grains with sizes lower than \unit[5]{nm} and can have a large impact on the destruction rate in intermediate- and high-density environments (up to 36\%).

The lowest forward shock dust destruction in the first \unit[10]{kyr} is found in the low-density ($n=\unit[1]{cm^{-3}}$) and strongly turbulent ($\mathcal{M}=3$) environment, where only 27\% of carbonaceous dust is destroyed. With \unit[0.85]{M$_\odot$} of carbonaceous dust, the least dust mass is destroyed in the $n=\unit[1]{cm^{-3}}$ and $\mathcal{M}=0.3$ case. However, the forward shock still has a high velocity of $\unit[v>250]{km\,s^{-1}}$ and is still hot at the end of these simulations, so that especially kinetic and thermal sputtering will process much more dust beyond the considered period. Compared to the average dust mass of \unit[0.1--1]{M$_\odot$} that is observed to be produced by SNRs and that still needs to survive the high-velocity reverse shock, we conclude that SNRs are rather net dust destroyers, in particular if higher gas densities are considered. 

If grain growth in the ISM were to be efficient, the total grain surface area would become more important than the destroyed dust mass. The many small grains that are produced by grain-grain collisions in our simulations could then even dominate the surface area of all grains and thus enable more efficient grain growth in the ISM, as discussed in Appendix~\ref{sec:appendix_number_grains}.

\section{Discussion}\label{sec:discussion}
Our study provides a comprehensive overview of the dust destruction efficiency by the forward shock during the first \unit[10]{kyr} of SNR evolution. It is the first to examine dust destruction in a turbulent medium that is specifically modeled to replicate ISM observations and has one of the highest resolutions of forward shock dust destruction studies because it focuses on the critical early stages of SNR evolution. A comparison to other studies is very valuable for this paper as they consider longer evolution times -- often even up to \unit[1]{Myr}.

\subsection{Previous studies}
By comparing our results to previous studies with longer evolution times, we can roughly estimate the amount of dust that would be destroyed by the forward shock if our simulations were longer, since we already concluded that most likely more dust will be destroyed beyond the considered first \unit[10]{kyr} of SNR evolution. Additionally, this comparison lets us assess how well previous models predicted their first few thousand years of dust destruction because they did not focus on this most destructive period. Usually, a lower spatial (and often temporal) resolution was used and some dust destruction mechanisms were often neglected. Table~\ref{tab:compare_studies} compares the main parameters and results of our own and other works after \unit[10]{kyr} of SNR evolution.

\begin{table*}[hbtp!]
    \centering
    \caption{\label{tab:compare_studies}Comparison of this study to different forward shock dust destruction studies after \unit[10]{kyr}.}
    \setlength{\tabcolsep}{4pt}
    \begin{tabular}{ccccccc}
        paper & ISM inhomogeneities & $n$ & resolution & gg & GSD & dust destr. for $n\approx\unit[1]{cm^{-3}}$ at \unit[10]{kyr} \\\hline
        this work & turbulence & \unit[1--100]{cm$^{-3}$} & \unit[0.022--0.044]{pc} & yes & MRN & \unit[0.85--1.53]{M$_\odot$} (27.3--56.8\%) \\ 
        $[1]$ & clumps & \unit[0.3--10]{cm$^{-3}$} & \unit[0.19--0.5]{pc} & no & MRN & \unit[0.67--1.17]{M$_\odot$} (32--39\%)\\
        $[2]$ & turbulence & \unit[0.7]{cm$^{-3}$} & \unit[0.5]{pc} & yes & MRN & \unit[2.8]{M$_\odot$}\\
         $[3]$ & weakly inhomogeneous & \unit[0.1--1]{cm$^{-3}$} & \unit[0.5]{pc} & yes & MRN & \unit[6.95]{M$_\odot$}\\
         $[4]$ & wind-driven bubble & \unit[1, 1000]{cm$^{-3}$} & \unit[0.11--0.21]{pc} & no & log-normal & \unit[0.40]{M$_\odot$}\\        
    \end{tabular}
    \setlength{\tabcolsep}{6pt}
    \tablefoot{[1] \citet{Dedikov2025}, [2] \citet{Kirchschlager2024a}, [3] \citet{Kirchschlager2022}, [4] \citet{Martinez-Gonzalez2019}. Compared are the ISM inhomogeneity model, the average ISM density ($n$), the resolution, whether grain-grain collisions are considered or not (gg), the initial grain-size distribution (GSD), and the destroyed dust mass (and rate) after the first \unit[10]{kyr} of SNR evolution for $n\approx\unit[1]{cm^{-3}}$ cases.}
\end{table*}

\citet{Dedikov2025} conduct such a study, using a similar approach to ours by simulating the forward shock dust destruction over \unit[100]{kyr} in a clumpy instead of turbulent medium, which is generated by fractal cubes. They use different inhomogeneity strengths and their densest clumps have roughly the same density contrasts between high- and low-density regions as our supersonic Mach number simulations. They do not consider magnetic fields nor grain-grain collisions, only kinetic and thermal sputtering. In their paper, they conclude that dense clumps can shield the dust efficiently, resulting in up to $\sim$20\% less destruction over \unit[100]{kyr} for an ISM with $n=\unit[1]{cm^{-3}}$ on average. After \unit[10]{kyr}, they find \unit[1.17]{M$_\odot$} ($\sim$39\%) 
of destroyed dust when considering a homogeneous medium, decreasing with increasing inhomogeneity down to \unit[0.67]{M$_\odot$} ($\sim$32\%)
of destroyed dust with the densest clumps so that these clumps already shield $\sim$7\% of the dust at \unit[10]{kyr}. Their considered grain species is not mentioned but the results match reasonably well to our carbonaceous (\unit[0.79--0.91]{M$_\odot$} or 25--33\%) grain simulations with $n=\unit[1]{cm^{-3}}$ that include only sputtering, for which we also find significant filament shielding of 8\% at \unit[10]{kyr}. In an intermediate-density environment ($\unit[n=10]{cm^{-3}}$) with rather dense ISM clumps, \unit[$\sim$4.5]{M$_\odot$} ($\sim$50\%)
of dust is destroyed in the first \unit[10]{kyr}, also fitting reasonably well to our results of the intermediate-density case with strong turbulence and carbonaceous grains (\unit[3.5--4.2]{M$_\odot$} or 46--63\% of dust destroyed). This supports the conclusion that generally more dust is destroyed at higher ISM densities. At \unit[100]{kyr}, however, this trend can reverse according to their study where \unit[8.0--11.4]{M$_\odot$} 
dust is destroyed in the low-density case, but only \unit[7.5]{M$_\odot$}
in the inhomogeneous intermediate-density case. 

If our sputtering results for carbonaceous grains would also align quite well with the results of \citet{Dedikov2025} beyond \unit[10]{kyr}, over \unit[8]{M$_\odot$} of dust would be destroyed at \unit[100]{kyr}, which is almost an order of magnitude more than what SNRs are observed to produce. Nevertheless, we note that the filaments in our study could behave differently than the clumpy medium of \citet{Dedikov2025} at lower shock velocities, and that we cannot estimate the dust evolution in our own simulations that include grain-grain collisions beyond \unit[10]{kyr} based on this comparison, as this destruction mechanism could become more efficient at later SNR stages (see for example our $\unit[n=100]{cm^{-3}}$ case).

\citet{Kirchschlager2024a} also investigate single SNRs in a turbulent ISM, similarly to us. However, they use SN-driven turbulence, which brings the caveats mentioned in Sect.~\ref{sec:intro} with it. After turbulence driving, they inject an SN in either a low- or intermediate-density area of \unit[$n=0.03 \text{ or } 0.7$]{cm$^{-3}$}, respectively. Thus, although their average ISM density is always lower than our lowest density case (\unit[$n=1$]{cm$^{-3}$}), the SN injection in their moderately dense region might be comparable to our simulations because their SNR does not encounter many voids, whereas we set off the SN in a void. Their dust properties and considered dust destruction mechanisms are similar to our study, but they only consider silicate grains.

After \unit[10]{kyr}, they find a destroyed dust mass of \unit[2.8]{M$_\odot$} in the intermediate-density simulation, which is almost two times more than the destroyed silicate grain mass in our high Mach turbulence study (\unit[1.5]{M$_\odot$}). This disparity can come from multiple factors. Most importantly, the turbulence driving is different such that the filaments they produce appear much wider than ours. Furthermore, the average ISM density in the box is lower with the SN set off in a higher density environment. Thus, the dust destruction efficiency can change significantly \citep{Kirchschlager2024a} and the encountered dust mass can be different. Moreover, their simulation is not fine-tuned to the first \unit[10]{kyr}, meaning that the temporal and spatial resolution is rather coarse (\unit[250]{yr} and \unit[0.5]{pc}) compared to our study (logarithmic time steps between \unit[0.15--120]{yr} depending on the evolution time and \unit[0.044]{pc} in the \unit[$n=1$]{cm$^{-3}$} cases), which can cause inaccuracies in the gas drag. With a lower spatial resolution and thus bigger grid cells, the dust velocity in a single cell is averaged over more grains, resulting in larger deviations compared to the real grain velocities. If the temporal resolution is too low, grains that are dragged with the forward shock may not be as efficiently decelerated by encountered ISM filaments as the gas so that the grains could be able to overtake the shock. This would leave the high-velocity grains exposed in a rather stationary ISM, leading to high relative gas-grain and grain-grain velocities. 

Thus, the higher destruction efficiency of \citet{Kirchschlager2024a} could be explained by an insufficient resolution. Other differences can add to these inaccuracies. First of all, the extrapolation from the 2D to the 3D dust mass is different and induces uncertainties. We multiply the averaged destruction efficiency of the three investigated 2D planes by the 3D SNR-encountered dust mass, whereas \citet{Kirchschlager2024a} fully extrapolate the dust mass from one 2D plane, using a correction factor they derive from comparing the SNRs in a homogeneous and an inhomogeneous medium. Thus, the real 3D destruction could be different. Secondly, they do not include partial vaporization, which was implemented in \texttt{Paperboats} by \citet{Kirchschlager2023}. This slightly reduces the dust destruction for the largest grains in our simulations. Thirdly, the penetration depth of gas ions into dust grains was calculated considering oxygen ions in their study, but was updated in our study to consider hydrogen instead of oxygen ions, which have a different penetration depth. This value alters the size-dependent sputtering, so that significantly more small silicate grains are destroyed. Lastly, it should be mentioned that the dust destruction in a turbulent ISM also depends on the detailed ISM environment, as we show with the three different orthogonal planes in this paper. The differences in the destruction efficiencies could therefore also partly come from the different surrounding environment and positions of the filaments. 

Similarly to our study, however, the derived dust destruction efficiencies of \cite{Kirchschlager2024a} do not plateau after \unit[10]{kyr}, but only start to plateau after \unit[50]{kyr} in their study. Nevertheless, they find that some dust continues to be destroyed even when their simulations finish at \unit[1]{Myr}, with a total of \unit[37]{M$_\odot$} of destroyed dust, which is $\sim$13.5 times more than what they find at \unit[10]{kyr}. If this trend is applied to our high Mach number and low-density ISM silicate simulation, it would mean that even in one of the lowest dust destruction cases, the SNR would destroy $\unit[1.5]{M_\odot}\times13.5\approx\unit[20]{M_\odot}$ of dust after \unit[1]{Myr}, being 20 times higher than the observed dust production in SNRs. This implies, once again, that SNRs are rather net dust destroyers.

The study of \citet{Kirchschlager2022} is similar to \citet{Kirchschlager2024a}, but only for a homogeneous medium and gas that is perturbed by weak Vishniac–Ostriker–Bertschinger overstabilities, which does not aim to resemble ISM turbulence. This medium can be best compared to our homogeneous and subsonic cases. They use the same setup and resolution as \citet{Kirchschlager2024a}, but consider an $\unit[n=1]{cm^{-3}}$ case, which is more comparable to our study. They find a destroyed dust mass of \unit[7.0]{M$_\odot$} in the first \unit[10]{kyr} which is reduced to \unit[1.3]{M$_\odot$} if only sputtering is considered, consistent across their inhomogeneous and homogeneous cases, which show almost identical destroyed dust masses. Their resulting dust destruction in the first SNR stages is, similarly to \citet{Kirchschlager2024a}, much higher than our results for this low-density environment, where we find $\sim$5 times less destroyed dust mass in the homogeneous and subsonic cases (\unit[1.4--1.5]{M$_\odot$}) but roughly the same destroyed dust masses with only sputtering considered (\unit[1.3--1.4]{M$_\odot$}). In addition to the differences mentioned in the comparison between our study and that of \citet{Kirchschlager2024a}, the study of \citet{Kirchschlager2022} also does not include a magnetic field, resulting in higher dust destruction due to weaker gas-grain coupling, as \citet{Kirchschlager2024a} discuss. Given these disparities between our own study and that of \citet{Kirchschlager2022}, the similarity in the sputtering-only simulations is most likely coincidental. Their simulations finish at \unit[1]{Myr} with $\unit[\sim$70$]{M_\odot}$ of destroyed dust, which is $\sim$10 times more than at \unit[10]{kyr}. If the SNRs in our study would follow the same trend, they would therefore again always destroy much more dust after \unit[1]{Myr} than they had originally produced.

Another useful comparison can be made to works that consider pre-SN feedback. \citet{Martinez-Gonzalez2019} let a wind-driven bubble evolve in a homogeneous ISM with a density of \unit[1]{cm$^{-3}$} before they inject the SN. The bubble's shocked region is compressed and has a density that is roughly three to four orders of magnitude higher than the interior of the bubble in which the SN explodes, resulting in roughly the same difference between filaments and voids as in our low-density and supersonic turbulence simulations. It expands up to \unit[23]{pc} in radius, meaning that it sweeps up \unit[$\sim$12.6]{M$_\odot$} of dust during the stellar wind evolution. They find that $\sim$\unit[0.36]{M$_\odot$} of dust (2.9\% of the shell's dust mass)
is destroyed in the first \unit[10]{kyr} after SN explosion if only thermal sputtering is considered and a mixture of silicate and carbonaceous grains is used. Although we also find that dust can be shielded by filaments, significantly more dust is shielded by their stellar wind shell than in our sputtering-only results with high Mach turbulence for either grain species (\unit[0.88--1.5]{M$_\odot$} and 25--45\%). However, their bubble is larger and its shell is  roughly \unit[3--5]{pc} much thicker than our filaments, meaning that it can slow down the shock more efficiently. As a result, most of the dust in the shell has not yet been encountered after \unit[10]{kyr} so that this dust cannot be destroyed. Furthermore, they do not include kinetic sputtering, which could be efficient and increase the destroyed dust fraction when the SNR forward shock starts to encounter the wind-driven bubble. At this point, grain-grain collisions could also become efficient due to the high dust densities and velocities, which would further increase the dust destruction.

Other studies that investigate the dust destruction of SNR forward shocks in a slightly inhomogeneous or turbulent ISM are \citet{Vasiliev2024} and \citet{Hu2019}, respectively. The former uses the same code as \citet{Dedikov2025} but only introduces small density perturbations of 10\% and does not compare their results to a homogeneous case so that the filament shielding efficiency cannot be assessed. Since they only study a monosized dust population, we cannot easily compare their study with ours because we use an MRN-distributed dust population, which leads to significantly different dust behavior. \citet{Hu2019} drive their turbulence with SNe, similar to \citet{Kirchschlager2024a}. However, they study the lifetime of silicate and carbonaceous dust grains over a total of \unit[400]{Myr} and several SNe. Thus, their destruction efficiencies are also not easily comparable to our study that only covers the first \unit[10]{kyr}. However, similarly to us, they find that silicates are better destroyed than carbonaceous grains.

\subsection{Caveats}
Although we include many mechanisms that are often neglected by other dust destruction studies, some uncertainties remain:
\begin{itemize}
    \item The pre-SN effects of stellar winds and radiation were only mimicked simplistically in our study by setting off the SNe in voids because a full pre-SN feedback model is out of the scope of this paper. In particular the weakly turbulent and homogeneous runs did not incorporate realistic pre-SN structures. For a forthcoming model, we plan to improve this by allowing stellar winds and radiation to evolve naturally in separate simulations, and letting the SN explode inside of the wind-driven bubble.
    
    \item Although we consistently set off all SNe in voids, the detailed surrounding ISM structures are not the same, since the turbulence-driving results in different filamentary shapes for each simulation. This affects the 3D SNR evolution in the first \unit[10]{kyr} and therefore also the dust destruction, resulting in an uncertainty, which could be quantified best by studying many different 3D ISM environments for every parameter set. Since the needed resources for this are too high, we assess this uncertainty with our investigation of the three orthogonal 2D planes, as discussed in Appendix~\ref{sec:appendix_diff_slices}. We find a maximum uncertainty of 9.4\% and include the standard deviation between those planes in our dust destruction results of Tables~\ref{tab:all_sims} and \ref{tab:all_sims_same_shock_vel}. This implies that the SNR dust destruction cannot be precisely predicted based on the initial parameters like ISM density and turbulent Mach number, but needs to take the exact SN ejection location (see also Appendix~\ref{sec:appendix_SN_at_filament}) and the surrounding environment structures into account.

    \item Our post-processing approach does not allow the dust to feed heavy elements coming from destroyed dust grains back to the gas, which physically should happen and would induce for example a faster cooling of the gas. The exact impact of the dust feedback is so far unknown but might be significant. Since on-the-fly approaches are more expensive than post-processing simulations, we used the latter for our calculations to include a more extensive dust destruction model. However, this brings the drawbacks of no dust-to-gas feedback and coarser time steps, which gives inaccuracies in the destruction, transport, and charge processes that we tried to minimize by taking as many logarithmically spaced snapshots as computationally feasible. A comprehensive on-the-fly approach is currently being implemented, tested, and compared against our post-processing method by our group (Sartorio et al., in prep.). Various on-the-fly models were used by other groups before \citep[e.g.,][]{Slavin2015, Martinez-Gonzalez2019, Vasiliev2024, Dedikov2025}; however, their dust destruction models were not as complete as in our post-processing code, so we chose \texttt{Paperboats}.

    \item We did not model the ejecta-dominated phase of the SNR evolution. Instead, we assumed that the ejecta mass can always be neglected in comparison to the encountered ISM mass and we inserted only thermal energy and no ejecta mass. Thus, the SNR evolution started immediately in the Sedov-Taylor phase instead of the ejecta-dominated phase. Furthermore, the pressure support of the ejecta was not modeled in this stage. The maximum amount of dust that would be encountered during the ejecta-dominated phase is 2--6\% of the encountered mass after \unit[10]{kyr} in our low ISM density simulations, depending on the considered ejecta mass (\unit[5--15]{M$_\odot$}). For our high- and intermediate-density ISM simulations, this fraction is $\sim$0.5--2.5\% and thus not significant. At low densities, the ejecta mass would be significant only if the ejecta-dominated phase were skipped entirely. Instead, we replaced it by an immediate Sedov-Taylor phase, which has a similar evolution to the late ejecta-dominated phase. Thus, only a fraction of the 2--6\% of the total dust mass is uncertain due to the missing ejecta mass and the dust destruction in our study can be expected to be only slightly underestimated in our low ISM density cases. However, since we started all of our simulations with an immediate Sedov-Taylor phase, concluded trends would not change qualitatively by considering ejecta mass.

    \item Because we did not include ejecta mass, we did not aim to resemble any specific SNR or SN type. Instead, we modeled SN type IIP and IIL in general by using a thermal energy of \unit[$10^{51}$]{erg} and having no pre-SN mass loss. Hence, our model only applies to this general approach and no observed SN.

    \item Chemistry was not fully traced. With an assumed collisional ionization equilibrium and an electron-ion equilibration, cooling times, dust charges, and dust transport could be influenced. Non-equilibrium ionization could for example decrease the temperatures at the shock front \citep{Slavin2015}, resulting in less efficient thermal sputtering. However, tracing ionization states in detail, and ion and electron temperatures separately, is beyond the scope of this study.

    \item Although \texttt{Arepo} uses time step subcycling to handle unresolved cooling times $t_\text{cool}$ \citep{Glover2007}, the cooling length was not fully resolved during the whole simulation. The cooling length, $l_\text{cool} = c_\text{s}t_\text{cool}$ with the speed of sound $c_\text{s}$, can be as short as \unit[$3\times10^{-7}$]{pc} in the shell of our high ISM density simulations because the ejecta cooling function peaks at \unit[$\sim$$2\times10^5$]{K}. This results in undercooled gas at the shock front, overestimating gas temperatures and therefore thermal sputtering. This undercooling seems to be a general problem of SNR dust destruction studies due to the large difference of the cooling length and the used numerical resolution. Thus, the significance of the unresolved cooling length cannot be easily assessed.
\end{itemize}

We emphasize that, while several assumptions may influence the absolute values of dust destruction in different ways, our model was applied consistently across all simulations. Therefore, the relative trends identified between simulations remain robust and meaningful.

\section{Conclusion}\label{sec:conclusion}
We performed 3D MHD \texttt{Arepo} simulations of SNRs evolving in a turbulent ISM that was driven to resemble observations more closely than in previous studies. Different turbulent ISM environments were considered with average ISM densities of $n=\unit[1,10, \text{ and } 100]{cm^{-3}}$ and turbulent Mach numbers of $\mathcal{M}=0,0.3,1, \text{ and } 3$. We let SNRs evolve for the most destructive first \unit[10]{kyr} and post-processed these simulations with \texttt{Paperboats} to study the dust destruction by the SNR forward shock, considering either purely carbonaceous or purely silicate dust. In detail, we find the following:
\begin{enumerate}
    \item The dust destruction by SNR forward shocks in their first \unit[10]{kyr} of evolution varies significantly, depending on the ISM environment. For \unit[$n=1-100$]{cm$^{-3}$} and $\mathcal{M}=0-3$, the destruction ranges from 27--92\%, which refers to \unit[0.85--11]{M$_\odot$} of dust. The largest dust mass is destroyed in the high-density environment. The least dust is destroyed in the low-density ISM where the shock still has end velocities of \unit[250--400]{km\,s$^{-1}$}. Since the total destroyed dust mass has not plateaued yet, it can be expected that more dust will be destroyed in the following years of SNR evolution, as is also derived from comparisons to other studies. Thus, we find that SNRs destroy more ISM dust than they produce, and thus are net dust destroyers.
    
    \item Filaments that closely resemble observations slow down the forward shock, and therefore shield some dust from destruction, similarly to the conclusions of other studies with different inhomogeneities or turbulence driving methods. While filaments in the low-density ISM only shield 8--14\% of the dust in the first \unit[10]{kyr} of SNR evolution, the intermediate-density cases result in 31--34\% less destruction in the highest Mach number simulations compared to the homogeneous ISM. In the high-density medium, the forward shock evolves into a thin and dense shell so that grain-grain collisions become significantly more important, resulting in a lower shielding ability of 11--14\%.
    
    \item Silicates are almost always more efficiently destroyed than carbonaceous grains, which is primarily caused by the higher sputtering yields. In a high-density ISM, the destruction difference between these two species decreases, partly caused by more effective grain-grain collisions, which are more efficient for carbonaceous grains than for silicates. In the weakly turbulent and homogeneous high-density cases, silicates are even more robust than carbonaceous grains.

    \item The dust destruction rate can vary even for similar turbulent environments with the same statistical parameters (average ISM density and turbulent Mach number) because the detailed structure of the turbulence changes, which affects the evolution time and shock velocity at which the first filaments are encountered. We can estimate this difference from the investigated three orthogonal planes, which show a difference in dust destruction rate of up to 9.4\% in the first \unit[10]{kyr}.
\end{enumerate}

Altogether, filaments in the turbulent ISM cannot shield dust enough for SNRs to act as net dust producers. Future work will incorporate pre-SN feedback and dust cooling into the workflow, which may further reduce dust destruction.

\begin{acknowledgements}
TS, FK, and IDL were supported by funding of the European Research Council (ERC) under the European Union’s Horizon 2020 research and innovation program DustOrigin (ERC-2019-StG-851622). TS and IDL acknowledge funding from the Flemish Fund for Scientific Research (FWO-Vlaanderen) through the research project G0A1523N. TS also acknowledges funding from FWO-Vlaanderen through the doctoral research project 11A5O26N. NSS acknowledges the support from the FWO-Vlaanderen in the form of a postdoctoral fellowship (1290123N). The resources and services used in this work were provided by the VSC (Flemish Supercomputer Center), funded by the FWO-Vlaanderen and the Flemish Government.
\end{acknowledgements}

\bibliographystyle{aa}
\bibliography{bib}

\end{nolinenumbers}
\begin{appendix}

\begin{nolinenumbers}
\section{SNR evolution}\label{sec:appendix_SNRevo}

  Figures~\ref{fig:gas_pics_ISM1}, \ref{fig:gas_pics_ISM10}, and \ref{fig:gas_pics_ISM100} show the temporal SNR evolution of all simulations in this study for the low-, intermediate-, and high-density cases, respectively. Their rows refer to the homogeneous, subsonic, transonic, and supersonic simulations and their columns to three different time steps. Links to movies of the SNR evolution are given in the captions of the respective evolution figures.

\begin{strip}
     \centering
     \begin{minipage}[t]{\linewidth}
     \centering
     \includegraphics[width=0.9\linewidth]{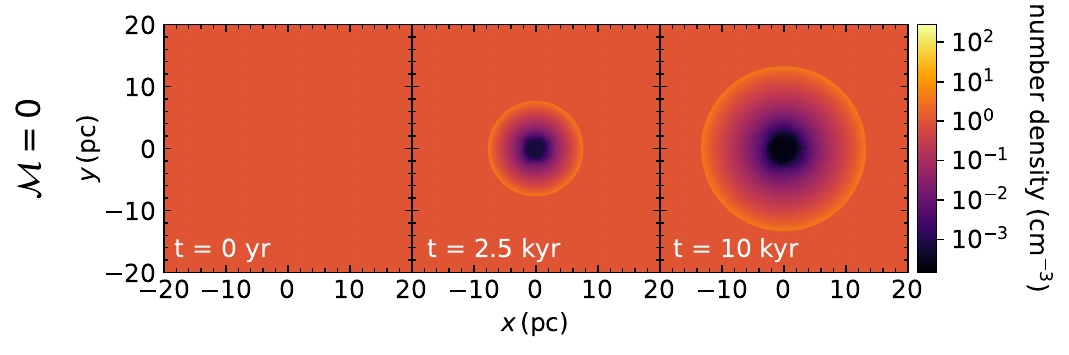}
     \end{minipage}
     \vspace{-20pt}

     \begin{minipage}[t]{\linewidth}
     \centering
     \includegraphics[width=0.9\linewidth]{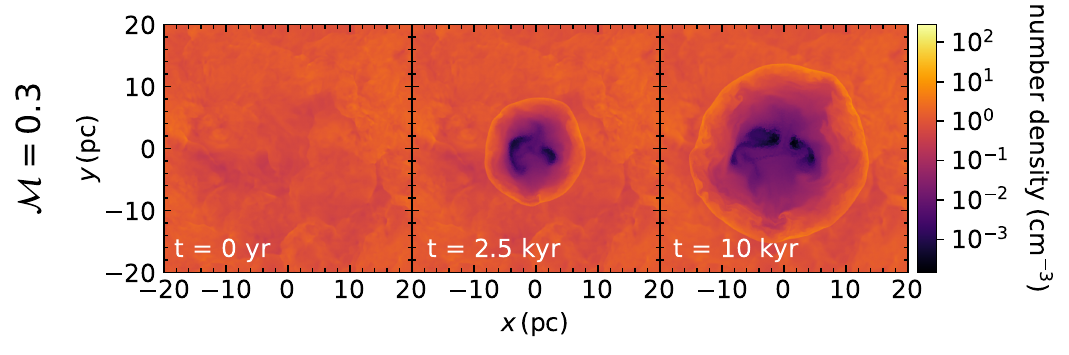}
     \end{minipage}
     \vspace{-20pt}

     \begin{minipage}[t]{\linewidth}
     \centering
     \includegraphics[width=0.9\linewidth]{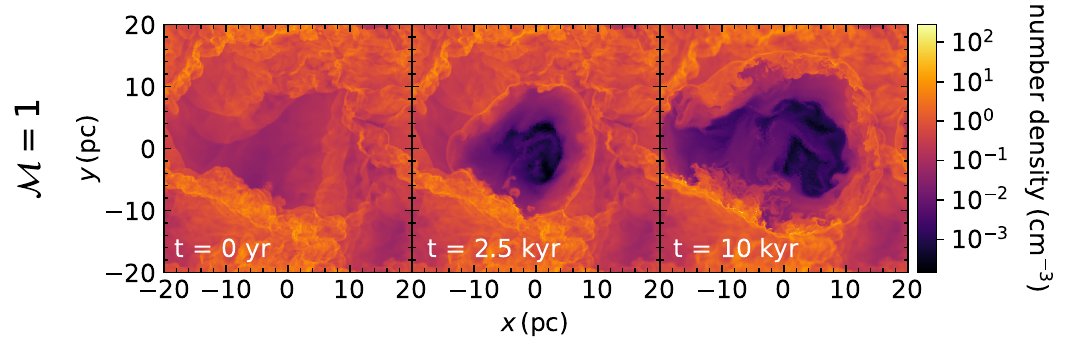}
     \end{minipage}
     \vspace{-20pt}

     \begin{minipage}[t]{\linewidth}
     \centering
     \includegraphics[width=0.9\linewidth]{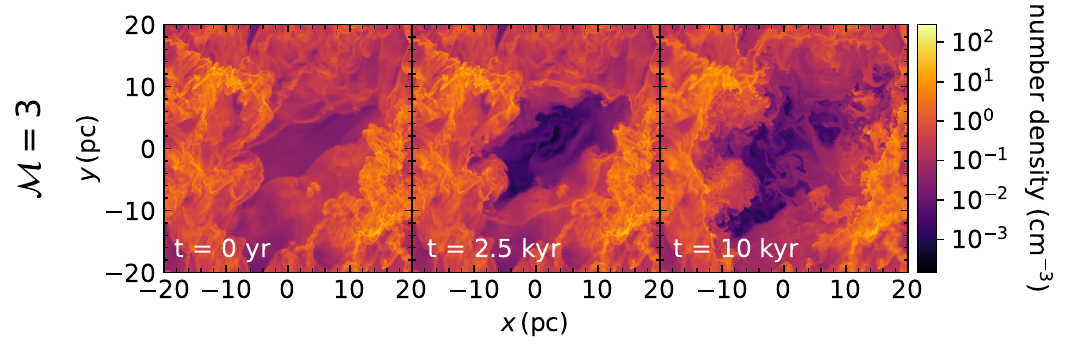}
     \end{minipage}
     \vspace{-21pt}
      \captionof{figure}{Number density of the $z=0$ slice before SN injection (left), during the SNR evolution at \unit[2.5]{kyr} (center), and at \unit[10]{kyr} (right) for an average ISM density of $\unit[n=1]{cm^{-3}}$. The rows show the different Mach numbers of $\mathcal{M}=0,0.3,1,\text{ and }3$ from top to bottom. The box size is always \unit[40]{pc} in each direction. Movies of these simulations can be found on \href{https://youtu.be/E2HJtMqN3wE}{https://youtu.be/E2HJtMqN3wE}.}
      \label{fig:gas_pics_ISM1}
\end{strip}

\FloatBarrier

\begin{figure*}[hbpt!]
     \centering
     \begin{minipage}[t]{\linewidth}
     \centering
     \includegraphics[page=1,width=.9\linewidth]{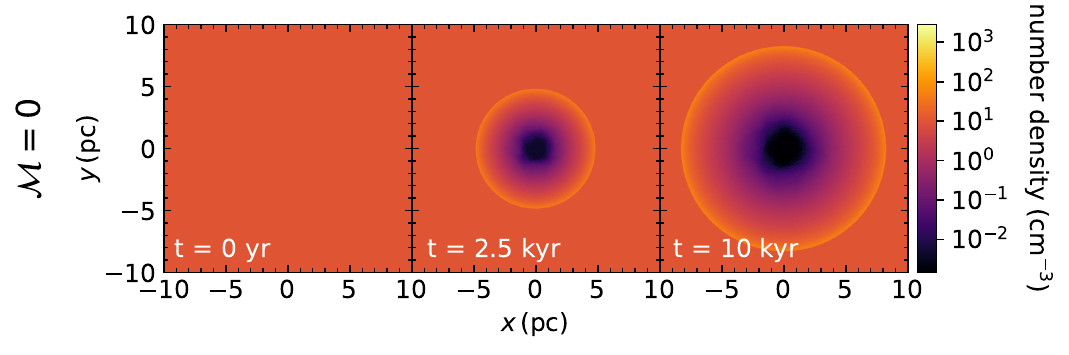}
     \end{minipage}
     \vspace{-20pt}

     \begin{minipage}[t]{\linewidth}
     \centering
     \includegraphics[page=1,width=.9\linewidth]{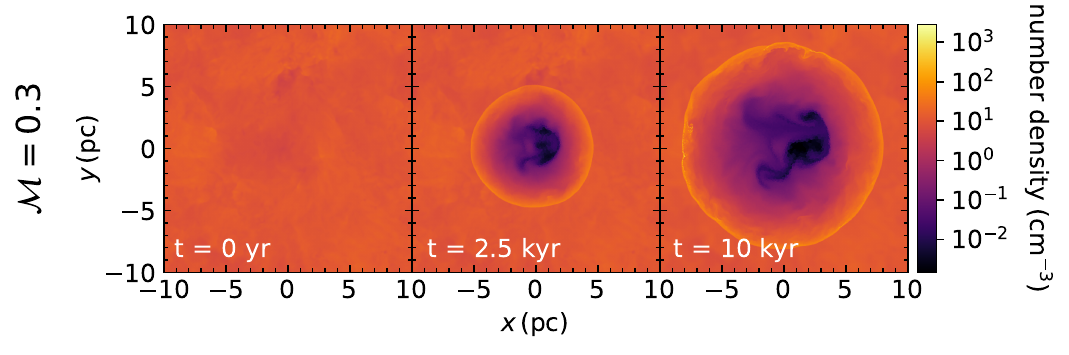}
     \end{minipage}
     \vspace{-20pt}

     \begin{minipage}[t]{\linewidth}
     \centering
     \includegraphics[page=1,width=.9\linewidth]{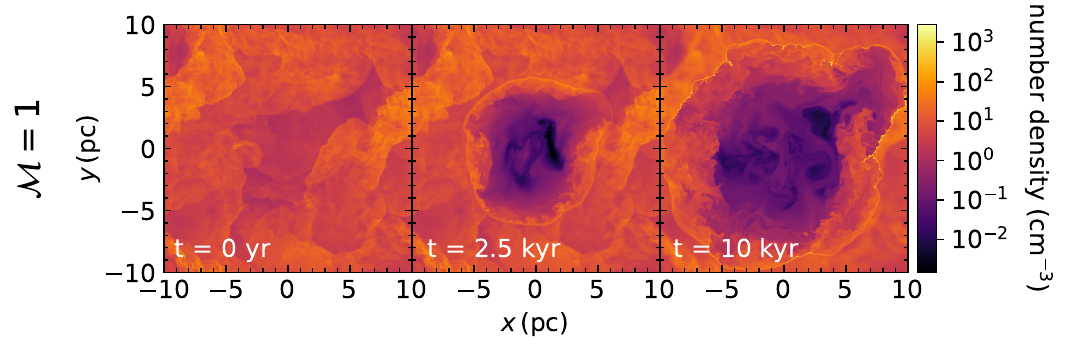}
     \end{minipage}
     \vspace{-20pt}

     \begin{minipage}[t]{\linewidth}
     \centering
     \includegraphics[page=1,width=.9\linewidth]{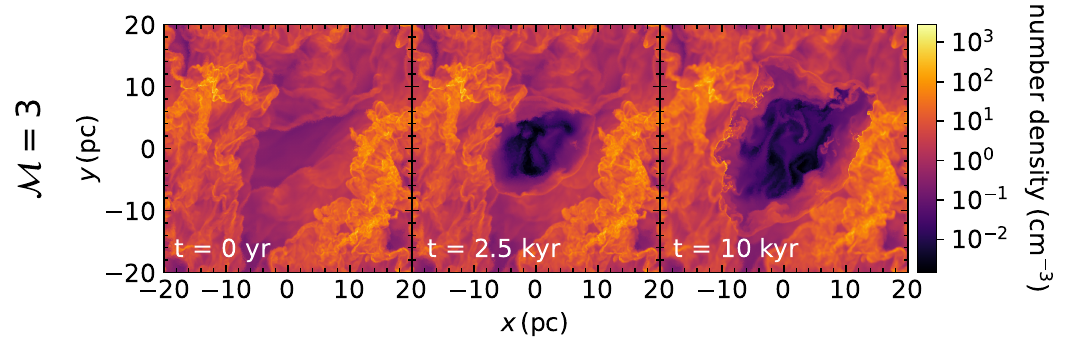}
     \end{minipage}
     \vspace{-20pt}
      \caption{Same as Fig.~\ref{fig:gas_pics_ISM1}, but for the $\unit[n=10]{cm^{-3}}$ cases. Here, only the $\mathcal{M}=3$ case has a box length of \unit[40]{pc}, whereas the other Mach number cases have a box length of \unit[20]{pc} in each direction. Movies of these simulations can be found on \href{https://youtu.be/boCK842iuds}{https://youtu.be/boCK842iuds}.}
      \label{fig:gas_pics_ISM10}
\end{figure*}

\FloatBarrier

\begin{strip}
     \centering
     \begin{minipage}[t]{\linewidth}
     \centering
     \includegraphics[page=1,width=.9\linewidth]{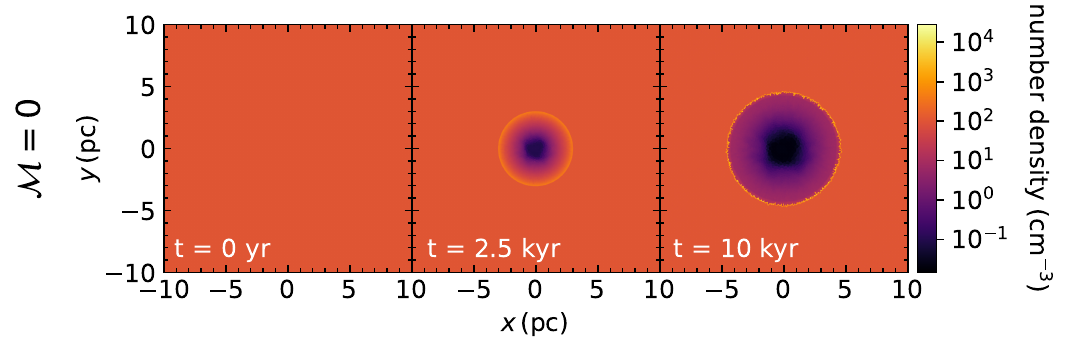}
     \end{minipage}
     \vspace{-20pt}

     \begin{minipage}[t]{\linewidth}
     \centering
     \includegraphics[page=1,width=.9\linewidth]{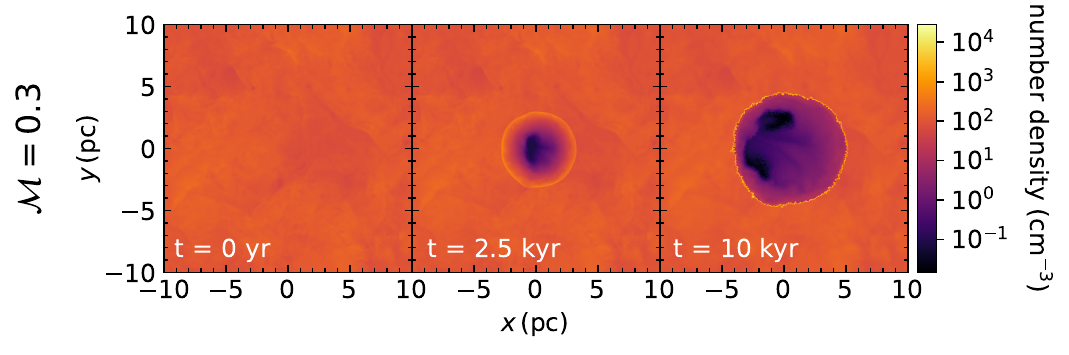}
     \end{minipage}
     \vspace{-20pt}

     \begin{minipage}[t]{\linewidth}
     \centering
     \includegraphics[page=1,width=.9\linewidth]{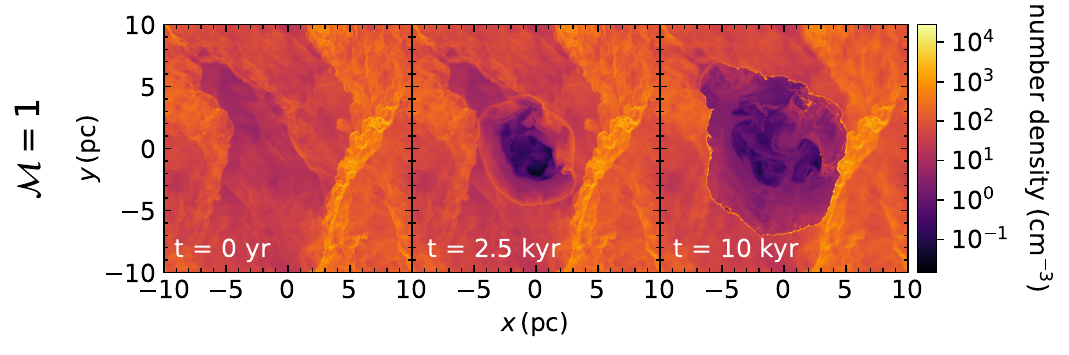}
     \end{minipage}
     \vspace{-20pt}

     \begin{minipage}[t]{\linewidth}
     \centering
     \includegraphics[page=1,width=.9\linewidth]{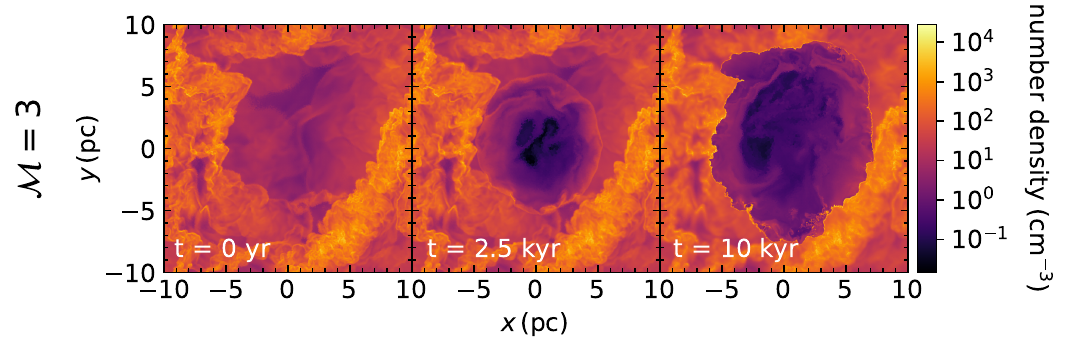}
     \end{minipage}
     \vspace{-20pt}
      \captionof{figure}{Same as Figs.~\ref{fig:gas_pics_ISM1} and \ref{fig:gas_pics_ISM10} but for the $\unit[n=100]{cm^{-3}}$ cases. Here, all boxes have a box length of \unit[20]{pc} in each coordinate direction. Movies of these simulations can be found on \href{https://youtu.be/FKQJa1xNakg}{https://youtu.be/FKQJa1xNakg}.}      
      \label{fig:gas_pics_ISM100}
  \end{strip}
  
\section{SN explosion close to filaments}\label{sec:appendix_SN_at_filament}
  \begin{figure*}[t!]
      \centering
      \includegraphics[width=.9\linewidth]{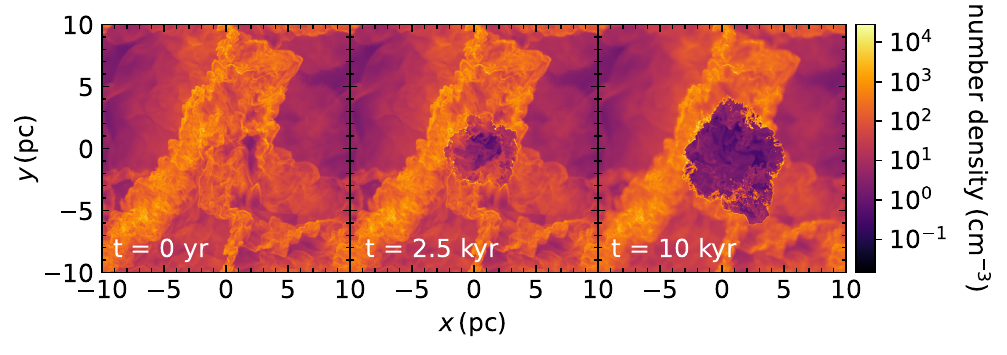}
      \caption{Number density of the $z=0$ slice before SN injection (left), during the SNR evolution at \unit[2.5]{kyr} (center), and at \unit[10]{kyr} (right) for a case in which the SN explodes close to filaments, and ISM parameters of $n=\unit[100]{cm^{-3}}$ and $\mathcal{M}=3$.}
      \label{fig:SN_at_filament}
  \end{figure*}
  
  \begin{table*}[hbtp!]
      \centering
      \caption{\label{tab:SN_at_filaments}Same as Table~\ref{tab:all_sims}, but for the simulation where the SN explodes close to filaments.}
      \begin{tabular}{cccccc}
           Si destr. (\%) & Si destr. (M$_\odot$) & C destr. (\%) & C destr. (M$_\odot$) & $v_\text{shock}^\text{end}$ & $t_\text{end}$ \\\hline
           86.2 $\pm$ 0.6  &  10.86 $\pm$ 0.08 & 85.5 $\pm$ 0.6  &  10.77 $\pm$ 0.08 & 89$\pm$3 & 10 \\ 
           87.0 $\pm$ 2.5  &  1.12 $\pm$ 0.03 & 79.3 $\pm$ 2.2  &  1.02 $\pm$ 0.03 & 387$\pm$13 & 1.07\\ 
           &\\
           \hline
           61.7 $\pm$ 3.3  &  7.77 $\pm$ 0.41 & 53.0 $\pm$ 2.6  &  6.68 $\pm$ 0.33 & 89$\pm$3 & 10 \\
           85.2 $\pm$ 3.2  &  1.09 $\pm$ 0.04 & 74.2 $\pm$ 3.3  &  0.95 $\pm$ 0.04 & 387$\pm$13 & 1.07
      \end{tabular}
    \tablefoot{This simulation considers an ISM density of $n=\unit[100]{cm^{-3}}$ and a turbulent Mach number of $\mathcal{M}=3$. The dust simulations with grain-grain collisions and sputtering (upper panel) are compared to the simulations that include only sputtering (lower panel). The dust destruction at the end of our simulations (\unit[10]{kyr}, first line) and at similar average shock velocities (\unit[$\sim$1]{kyr}, second line) to all other cases (Appendix~\ref{sec:appendix-same_FS_v}) are shown.}
  \end{table*}

\begin{figure*}[hbpt!]
     \centering
     \begin{minipage}[t]{.45\linewidth}
     \centering
     \includegraphics[width=.9\linewidth]{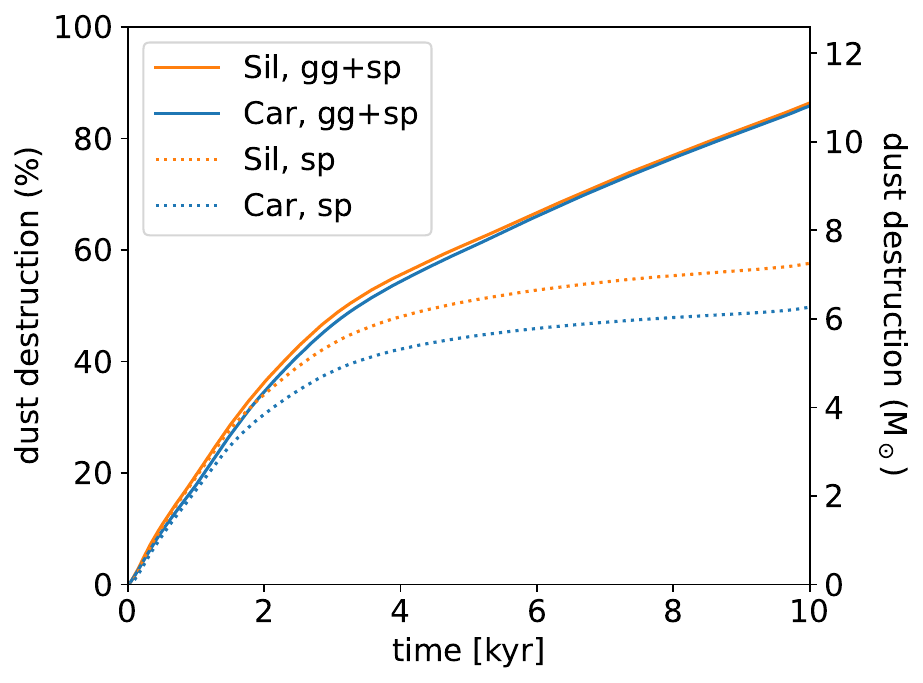}
     \end{minipage}
     \begin{minipage}[t]{.45\linewidth}
     \centering
     \includegraphics[width=.9\linewidth]{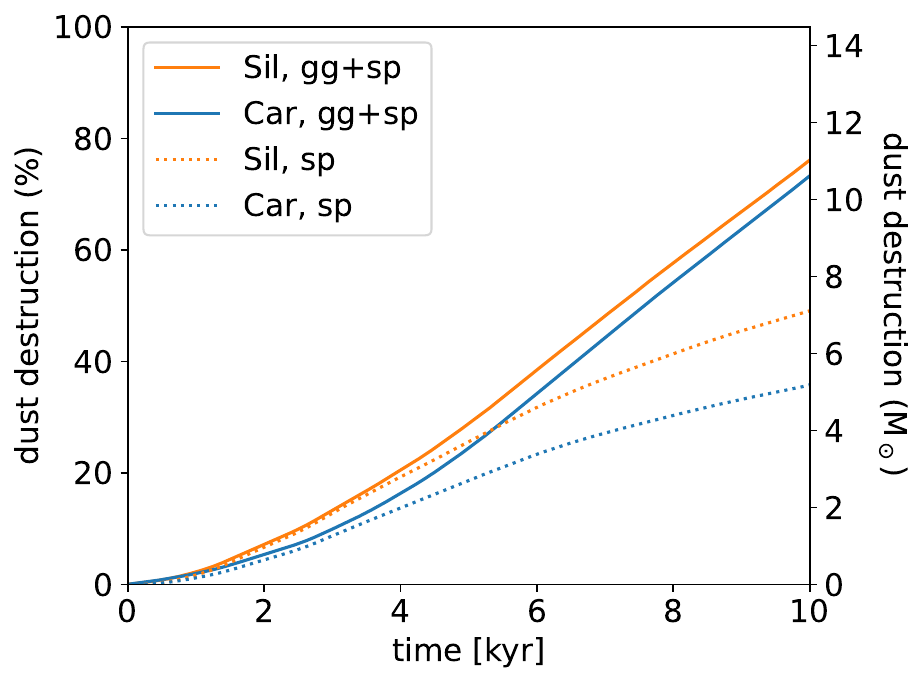}
     \end{minipage}
      \caption{Dust destruction of the $z=0$ slice for the case in which the SN explodes close to filaments (left) and for the case in which the SN explodes in a void (right, also shown in Fig.~\ref{fig:destr_results}). Both simulations have an ISM density of $n=\unit[100]{cm^{-3}}$ and Mach number of $\mathcal{M}=3$, and include either silicate (Sil) or carbonaceous (Car) grains, and grain-grain collisions with sputtering (gg+sp) or only sputtering (sp).}
      \label{fig:SN_at_filament_destr}
  \end{figure*}
  
 The position of the SN in the ISM has a significant impact on its dust destruction. To make all simulations comparable and to mimic pre-SN feedback, we always placed the SN in a large void of the driven turbulence. To quantify the impact of the explosion position, we ran a simulation of the $n=\unit[100]{cm^{-3}}$ and $\mathcal{M}=3$ case in which the SN explodes close to filaments, as shown in Fig.~\ref{fig:SN_at_filament}, and compared it to the fiducial $n=\unit[100]{cm^{-3}}$ and $\mathcal{M}=3$ case used in this paper. In this simulation, the SNR only expands up to \unit[$\sim$5]{pc} in radius, whereas it expands to \unit[$\sim$8]{pc} in the case where it evolves in a void (bottom panel of Fig.~\ref{fig:gas_pics_ISM100}). The average shock velocity at \unit[10]{kyr} (see Table~\ref{tab:SN_at_filaments}) is roughly the same between both cases. However, the SNR in a void sweeps up more gas during its evolution because it only encounters dense filaments at the end of the simulation, after it already expanded through a substantial amount of material. The dense filaments can then almost stop the shock since it was continuously slowed down by lower density regions at earlier times. The SNR close to filaments is slowed down early due to the high gas densities around it. At such early stage, its forward shock is energetic enough to penetrate the filaments and keep slowly expanding.

  This difference in SNR evolution affects the dust destruction efficiency. The SNR that explodes close to filaments encounters more dust in its early evolutionary stage where its forward shock is highly energetic. Thus, the destroyed dust mass increases rapidly at the beginning of the simulation, as shown in Fig.~\ref{fig:SN_at_filament_destr} but decreases toward the end where the shock is much slower and does not encounter as much dust anymore. The silicate and carbonaceous grains simulations show similar total dust destruction efficiencies of $\sim$86\% (Table~\ref{tab:SN_at_filaments}), which is more than 10\% more than for the  SN explosion in a void (Table~\ref{tab:all_sims}, 76\% and 73\% for the silicate and carbonaceous case, respectively). The total destroyed dust mass after \unit[10]{kyr} is similar, however, because the SNR in a void encounters more material.

  When the SNR evolves closer to filaments, it furthermore forms a shell at earlier times since radiative cooling quickly becomes effective due to the high masses of swept-up gas at early times. Therefore, grain-grain collisions become important early on (from \unit[$\sim$2--3]{kyr} onward), resulting in a final difference with the sputtering-only simulations of 25\% and 33\% for silicate and carbonaceous grains, respectively, as shown in Table~\ref{tab:SN_at_filaments}. Grain-grain collisions are also the main reason why carbonaceous grains are similarly well destroyed as silicates in this simulation.

  All in all, the dust destruction efficiency of the SNR close to filaments is higher than when it explodes in a void, but similar to the homogeneous and weakly turbulent $n=\unit[100]{cm^{-3}}$ cases after \unit[10]{kyr} of evolution (87\% for the silicate and carbonaceous grain simulations). Hence, the filaments cannot shield the dust from destruction at early times if the SN explodes closer to the dense filamentary structures of ISM turbulence. Instead, they then lead to one more solar mass of destroyed dust than in the SN in a homogeneous ISM. However, the SNR at filaments is also slowed down quicker than the homogeneous case SNR so that the destruction efficiency and destroyed dust mass are smaller than in the homogeneous case at similar end shock velocities (see Appendix~\ref{sec:appendix-same_FS_v}). After \unit[1]{Myr} of SNR evolution, the SNR close to filaments might therefore result in lower dust destruction than for the homogeneous high-density case.

  \section{Resolution of the dust simulations}\label{sec:appendix_4500cells} 
  Different Cartesian grid resolutions were investigated for the \texttt{Paperboats} simulations and $900\times 900 \times 1$ was found to be optimal for our setup as it did not introduce significant errors while keeping the computational time feasible.
  However, at the last few time steps ($\unit[\sim$9--10$]{kyr}$), this resolution is not sufficient anymore to resolve the forward shock, which gets thinner due to the expansion of the shock. Thus, the dust transport is not sufficiently resolved anymore, leading to dust grains that overtake the forward shock when filaments are encountered since the gas in the shock front is more efficiently stopped than the dust. We carefully investigated this phenomenon and found that it is not of physical origin but disappears with a higher considered resolution of $4500\times4500\times1$ cells, which would require on the order of 25 times more computational resources, which is beyond our capabilities. The maximum possible resolution with our setup was 1800 cells per box side, which we tested against the 900 cells simulation for the high-density supersonic Mach number simulation. With the higher resolution, the majority of the dust remains inside of the SNR; however, at some parts with an especially dense and thin shell, the dust still overtakes the forward shock. Comparing these two resolutions, the dust destruction only changes by 0.28\%, meaning that this phenomenon introduces a negligible error compared to other more significant uncertainties in this study. Since this phenomenon only occurs in the last few time steps and therefore only hardly changes the dust destruction over the full \unit[10]{kyr}, we chose to carry out the analysis with 900 cells per side, and we note that the last one thousand years should be evaluated more carefully for this study.

  \section{Influence of  the surrounding ISM structure}\label{sec:appendix_diff_slices}

  The SNe in this study were consistently injected in large void regions. The exact distance to nearby filaments and the detailed structure of the ambient ISM, however, still influence the dust destruction by a significant factor because they determine the velocity at which the forward shock encounters high-density and dust-rich regions. Our dust simulations with \texttt{Paperboats} only consider three orthogonal 2D slices of the 3D SNR simulations from which we then take the average value for our results. Since the turbulence in these slices has the same statistical properties and the injection region is always a void, the dust destruction difference between these slices can be used to estimate the influence of the detailed structures found in the ambient medium. The difference in dust destruction efficiency between these slices ranges from 0.01\% up to 9.4\% (Fig.~\ref{fig:slicediff}), proving that this can have a large impact, especially for the high Mach number simulations since the slice structure can differ more significantly there. The dust destruction difference in one snapshot can increase even more if the SNe are injected in regions with higher densities than voids, as shown by \citet{Kirchschlager2024a}.

  \begin{figure}[hbtp!]
      \centering
      \includegraphics[width=\linewidth]{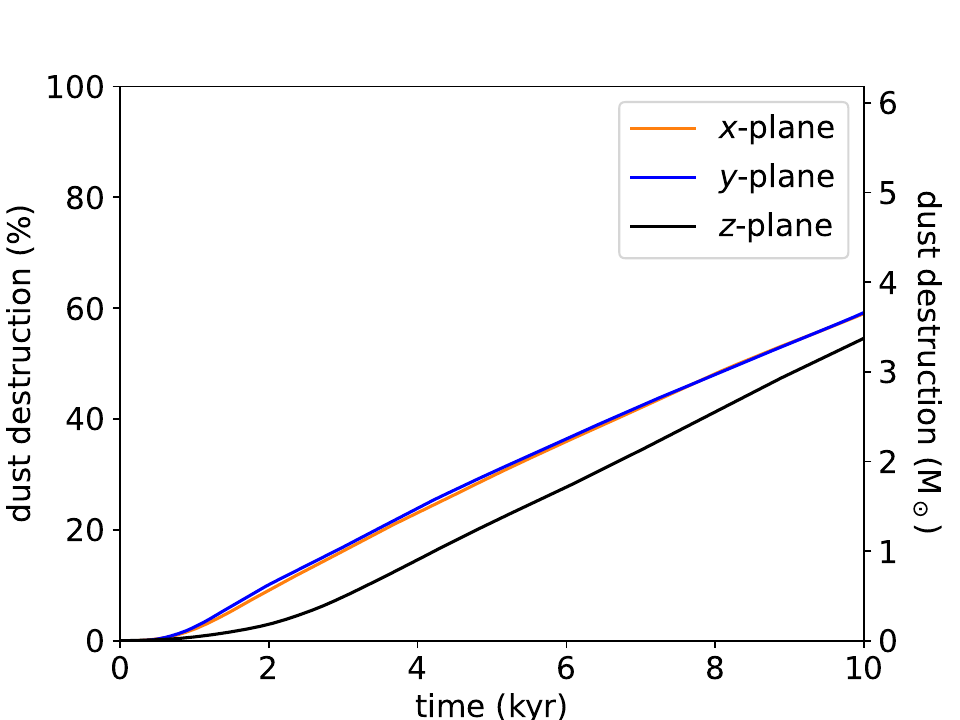}
      \caption{Difference in dust destruction between three orthogonal planes of the same 3D MHD simulation ($\unit[n=10]{cm^{-3}}$ and $\mathcal{M}=3$). Although the turbulence always has the same statistical properties, the dust destruction varies by roughly 9.4\%, depending on the exact density distribution and close-by filamentary shapes.}
      \label{fig:slicediff}
  \end{figure}

  \section{Grain-size distributions}\label{sec:appendix-GSDs}

  \begin{figure*}[hbtp!]
      \centering
      \includegraphics[width=\linewidth]{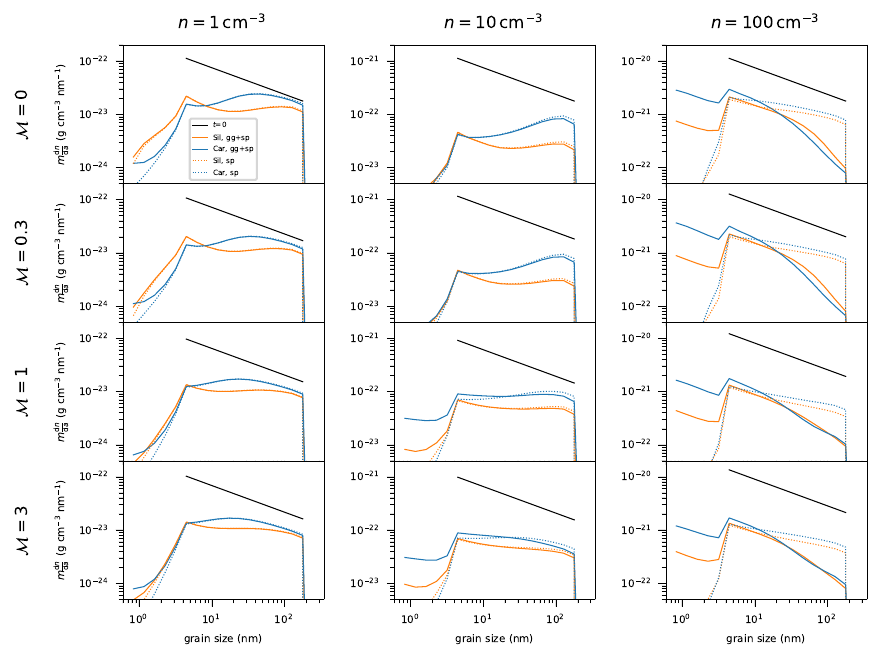}
      \caption{Grain-size distributions at \unit[10]{kyr} compared to the initial MRN distribution (black) for the carbonaceous and silicate grains. The $y$-axis represents the dust mass density per particle radius, which is the number of dust grains per volume and grain radius $\text{d}n/\text{d}a$ multiplied with the particle mass $m$. All $z=0$ simulation parameter sets are shown, including the different considered turbulent Mach numbers ($\mathcal{M}$) and average ISM densities ($n$). The plots differentiate between carbonaceous (blue) and silicate (orange) dust, and the grain-grain collisions and sputtering (solid), and the sputtering-only (dotted) simulations.}
      \label{fig:GSDs}
  \end{figure*}
  
  The grain-size distributions of all simulations are shown in Fig.~\ref{fig:GSDs}. The black lines show the initial MRN distribution for the carbonaceous and silicate dust, which is altered due to the destruction by the forward shock. The final distributions are similar for the different Mach number cases but differ significantly between the ISM densities. In a low-density environment, grain-grain collisions are inefficient. Thus, large grains above \unit[$\sim$40]{nm} mostly survive. However, a large fraction of intermediate-sized grains down to \unit[5]{nm} is sputtered so that the final distribution is altered. Grains that are smaller than \unit[5]{nm} are not included in the initial distribution but a significant number is generated by the different destruction mechanisms. The final grain-size distributions of the intermediate-density cases are similar but with more destroyed dust. Thus, the difference between initial and final distribution increases. Furthermore, the carbonaceous grain-size distributions show a variation between the different Mach numbers, as large carbonaceous grains are better destroyed in a strongly turbulent medium than in low Mach number turbulence. Only the high-density ISM simulations deviate significantly from this distribution. There, grain-grain collisions become important because of the formation of the thin and dense shell. Therefore, many large grains are fragmented to smaller sizes so that intermediate-sized grains are less efficiently destroyed than larger grains. In addition, many more grains smaller than \unit[5]{nm} are generated than in the lower density cases. However, if only sputtering is considered, the final grain-sizes are distributed similarly to the intermediate-density cases.

  \section{Dust destruction results with similar forward shock end velocities}\label{sec:appendix-same_FS_v}
  Table~\ref{tab:all_sims_same_shock_vel} shows the dust destruction results for all simulations of this paper, similar to Table~\ref{tab:all_sims}, but ending at the same forward shock velocity averaged over the whole forward shock, instead of the same evolution time. This is a useful comparison between the various Mach number and ISM density cases investigated because the SNR evolves to different evolutionary stages with different shock velocities in our different models. However, most conclusions gathered from Table~\ref{tab:all_sims} still hold true in this comparison so that we only attach it in the appendix. The most significant change is that silicates are now always more efficiently destroyed than carbonaceous grains and that the dust destruction increases with higher ISM densities for all simulations without exceptions. Furthermore, the destroyed dust masses decrease with higher Mach numbers in this comparison, contrary to Table~\ref{tab:all_sims} where they remain roughly the same. Other found trends stay consistent.
  
  \begin{table*}[hbtp!]
  \centering
  \caption{\label{tab:all_sims_same_shock_vel}Same as Table~\ref{tab:all_sims}, but for a similar average end shock velocity of $\unit[387-390]{km\,s^{-1}}$, and thus different end times $t_\text{end}$.}
  
  \begin{tabular}{cccccccc}
    $n$ (cm$^{-3}$) & $\mathcal{M}$ & Si destr. (\%) & Si destr. (M$_\odot$) & C destr. (\%) & C destr. (M$_\odot$) & $v_\text{shock}^\text{end}$ (km\,s$^{-1}$) &$t_\text{end}$ (kyr) \\\hline
    
    1   & 0 & 56.8  &  1.46 & 35.5  &  0.91 & 388$\pm$3 & 10.00 \\ 
    1   & 0.3& 50.1 $\pm$ 1.9  &  0.89 $\pm$ 0.03 & 31.4 $\pm$ 1.2  &  0.56 $\pm$ 0.02 & 389$\pm$2 & 8.53\\ 
    1   & 1 & 31.2 $\pm$ 3.1  &  0.31 $\pm$ 0.03 & 19.3 $\pm$ 1.7  &  0.19 $\pm$ 0..02 & 389$\pm$9 & 4.57\\ 
    1   & 3 & 36.5 $\pm$ 3.9  &  0.60 $\pm$ 0.06 & 22.6 $\pm$ 2.2  &  0.37 $\pm$ 0.04 & 390$\pm$20 & 5.84\\\\ 

    10  & 0 & 87.6  &  2.92 & 67.5  &  2.25 & 389$\pm$3 & 4.51\\ 
    10  & 0.3& 86.0 $\pm$ 0.2  &  1.62 $\pm$ 0.01 & 65.4 $\pm$ 0.3  &  1.23 $\pm$ 0.01 & 387$\pm$2 & 4.05\\ 
    10  & 1 & 74.4 $\pm$ 2.1  &  1.20 $\pm$ 0.03 & 52.8 $\pm$ 2.3  &  0.85 $\pm$ 0.04 & 389$\pm$7 & 3.58\\ 
    10  & 3 & 48.5 $\pm$ 5.2  &  0.82 $\pm$ 0.09 & 33.6 $\pm$ 4.8  &  0.57 $\pm$ 0.08 & 387$\pm$25 & 3.58\\\\ 
    
    100 & 0 & 95.5  &  3.67 & 90.3  &  3.48 & 389$\pm$1 & 2.00\\ 
    100 & 0.3& 95.0 $\pm$ 0.1  &  1.70 $\pm$ 0.01 & 89.0 $\pm$ 0.2  &  1.60 $\pm$ 0.01 & 387$\pm$3 & 1.84\\ 
    100 & 1 & 89.7 $\pm$ 0.6  &  1.74 $\pm$ 0.01 & 77.1 $\pm$ 2.0  &  1.50 $\pm$ 0.04 & 388$\pm$9 & 2.52\\ 
    100 & 3 & 71.6 $\pm$ 0.2  &  1.02 $\pm$ 0.01 & 54.0 $\pm$ 0.5  &  0.77 $\pm$ 0.01 & 390$\pm$7 & 2.55\\ 

    \\
    \hline

    1   & 0 & 55.5  &  1.42 & 33.0  &  0.85 & 388$\pm$3 & 10.00 \\ 
    1   & 0.3& 48.6 $\pm$ 1.9  &  0.86 $\pm$ 0.03 & 28.8 $\pm$ 1.0  &  0.51 $\pm$ 0.02 & 389$\pm$2 & 8.53\\ 
    1   & 1 & 30.3 $\pm$ 3.1  &  0.30 $\pm$ 0.03 & 17.6 $\pm$ 1.7  &  0.17 $\pm$ 0.02 & 389$\pm$9 & 4.57\\ 
    1   & 3 & 35.6 $\pm$ 3.9  &  0.58 $\pm$ 0.06 & 20.8 $\pm$ 2.3  &  0.34 $\pm$ 0.04 & 390$\pm$20 & 5.84\\\\ 

    10  & 0 & 86.0  &  2.87 & 62.7  &  2.09 & 389$\pm$3 & 4.51\\ 
    10  & 0.3& 84.2 $\pm$ 0.2  &  1.59 $\pm$ 0.01 & 60.3 $\pm$ 0.4  &  1.14 $\pm$ 0.01 & 387$\pm$2 & 4.05\\ 
    10  & 1 & 72.5 $\pm$ 2.1  &  1.17 $\pm$ 0.03 & 48.3 $\pm$ 2.3  &  0.78 $\pm$ 0.04 & 389$\pm$7 & 3.58\\ 
    10  & 3 & 46.3 $\pm$ 4.9  &  0.78 $\pm$ 0.08 & 29.2 $\pm$ 4.2  &  0.49 $\pm$ 0.07 & 387$\pm$25 & 3.58\\\\ 
    
    100 & 0 & 94.3  &  3.63 & 87.7  &  3.38 & 389$\pm$1 & 2.00\\
    100 & 0.3&93.9 $\pm$ 0.1  &  1.68 $\pm$ 0.01 & 86.0 $\pm$ 0.3  &  1.54 $\pm$ 0.01 & 387$\pm$3 & 1.84\\ 
    100 & 1 & 88.3 $\pm$ 0.8  &  1.71 $\pm$ 0.02 & 72.7 $\pm$ 2.5  &  1.41 $\pm$ 0.05 & 388$\pm$9 & 2.52\\ 
    100 & 3 & 68.1 $\pm$ 0.4  &  0.98 $\pm$ 0.01 & 47.0 $\pm$ 0.5  &  0.67 $\pm$ 0.01 & 390$\pm$7 & 2.55 
  \end{tabular}
  
\end{table*}
\FloatBarrier
  \section{Total surface area of dust grains}\label{sec:appendix_number_grains}
  If grain growth of dust grains in the ISM is found to be efficient, the total grain surface area in the post-shock region is an important parameter to evaluate as these grains could grow in the ISM once they are formed, dependent on their surface area. In all of our parameter sets, we observe that the total number of dust grains significantly increases because grain-grain collisions generate many small grains (see Appendix~\ref{sec:appendix-GSDs}). The total grain surface area, however, has a more complex evolution and can decrease if too much dust is destroyed but also increase in the case that grain-grain collisions are fragmenting enough larger dust grains to smaller sizes with a higher total grain surface area but sputtering is not efficient enough to destroy these grains. Thus, the ratio between the total grain surface area at the end to the initial value at the beginning is dependent on the considered parameter set, as can be seen in Table~\ref{tab:grain_surface_area}. The total grain surface area of silicate grains is always found to decrease; however, it decreases much less for the high-density simulations in which grain-grain collisions are efficient. For carbonaceous grains, the total grain surface area increases in the high-density cases and the intermediate-density case with a strongly turbulent ISM since grain-grain collisions are more efficient for carbonaceous grains than for silicates (see Sect.~\ref{subsubsec:SNR_low_dens}). All other parameter sets show a decrease of carbonaceous grain surface area after \unit[10]{kyr} compared to the initial dust distribution because grain-grain collisions are not that efficient in these simulations. Together with efficient grain growth in the ISM, this could therefore indicate a significantly lower destruction fraction of the SNR forward shock after the most destructive \unit[10]{kyr} in intermediate- and high-density cases where grain-grain collisions are important. For some environments, accretion could even be up to almost three times more efficient after the SNR shocks the ISM than before the SN event.

\begin{table}[hbtp!]
      \centering
      \caption{\label{tab:grain_surface_area}Temporal change of the total grain surface area within the SNR shocked region.}
      \rotatebox{90}{\hspace{-3.9em} Car \hspace{3.7em} Sil}  
      \begin{tabular}{cl|ccc}
          & & $n=\unit[1]{cm^{-3}}$ & $n=10$ & $n=100$ \\\hline
          & $\mathcal{M}=0$ & 0.43 & 0.08 & 0.77  \\
          & $\mathcal{M}=0.3$ & 0.43 $\pm$ 0.03 & 0.08 $\pm$ 0.01 & 0.84 $\pm$ 0.02\\
          &$\mathcal{M}=1$ & 0.27 $\pm$ 0.02 & 0.25 $\pm$ 0.03 & 0.71 $\pm$ 0.17 \\
          &$\mathcal{M}=3$ & 0.29 $\pm$ 0.01 & 0.5 $\pm$ 0.25 & 0.66 $\pm$ 0.14\\&\\

          &$\mathcal{M}=0$ & 0.35 & 0.10 & 2.42  \\
          &$\mathcal{M}=0.3$ & 0.39 $\pm$ 0.03 & 0.11 $\pm$ 0.01 & 2.92 $\pm$ 0.05 \\
          &$\mathcal{M}=1$ & 0.32 $\pm$ 0.02 & 0.71 $\pm$ 0.15 & 2.27 $\pm$ 0.57 \\
          &$\mathcal{M}=3$ & 0.36 $\pm$ 0.03 & 1.29 $\pm$ 0.72 & 1.73 $\pm$ 0.37\\
      \end{tabular}
    \tablefoot{The ratio of the total grain surface area at $t=\unit[10]{kyr}$ to its initial value at $t=\unit[0]{yr}$ is shown, depending on the considered average ISM number density ($n$), turbulent Mach number ($\mathcal{M}$), and grain species (Car for carbonaceous and Sil for silicate grains). A value higher than 1 means that the total grain surface area increased, whereas a value lower than 1 means that it decreased.}
  \end{table}  
  
\end{nolinenumbers}
\end{appendix}
\end{document}